\documentclass[sigconf,nonacm]{acmart}

\usepackage{tikz}
\usepackage{amsmath,amsfonts}
\usepackage[linesnumbered,vlined]{algorithm2e}
\usepackage{graphicx}
\usepackage{textcomp}
\usepackage{xcolor}
\usepackage{xspace}
\usepackage{tabularx}
\usepackage{makecell}
\usepackage{paralist}
\usepackage{caption}
\usepackage{subcaption}
\usepackage{soul}
\usepackage{mdframed}
\usepackage[inline,shortlabels]{enumitem}
\usepackage{balance}

\newcommand{\thesystem}{Destiny\xspace}
\newcommand{\theparadigm}{DQBFT\xspace}
\newcommand{\thesubsystem}{ThreshSign\xspace}

\makeatletter
\newcommand{\removelatexerror}{\let\@latex@error\@gobble}
\let\oldnl\nl
\newcommand{\nonl}{\renewcommand{\nl}{\let\nl\oldnl}}

\makeatother

\newcolumntype{L}[1]{>{\raggedright\arraybackslash}p{#1}}
\newcolumntype{C}[1]{>{\centering\arraybackslash}p{#1}}
\newcolumntype{R}[1]{>{\raggedleft\arraybackslash}p{#1}}

\newcommand*\circled[1]{\tikz[baseline=(char.base)]{
		\node[shape=circle,draw,inner sep=1pt,fill=green!20] (char) {#1};}}

\begin{document}
\title{Scalable Byzantine Fault Tolerance via Partial Decentralization}

\author{Balaji Arun}
\affiliation{%
  \institution{Virginia Tech}
}
\email{balajia@vt.edu}

\author{Binoy Ravindran}
\affiliation{%
  \institution{Virginia Tech}
}
\email{binoy@vt.edu}

\begin{abstract}
    Byzantine consensus is a critical component in many \emph{permissioned} 
    Blockchains and distributed ledgers.
    We propose a new paradigm for designing BFT protocols called \theparadigm 
    that addresses three major performance and scalability  challenges 
    that plague past protocols: 
    (i) high communication costs to reach geo-distributed agreement, 
    (ii) uneven resource utilization hampering performance,  and 
    (iii) performance degradation under varying node and network conditions 
    and high-contention workloads. Specifically, \theparadigm divides  
    consensus into two parts: 
    1) durable command replication without a global order, and 
    2) consistent global ordering of commands across all replicas. 
    \theparadigm achieves this by decentralizing the heavy task of replicating 
    commands while centralizing the ordering process. 
        
    Under the new paradigm, we develop a new protocol, 
    \thesystem that 
    uses 
     a combination of three techniques to achieve high performance 
    and scalability: using a 
    trusted subsystem to decrease consensus's quorum size, using  
    threshold signatures to attain linear communication costs, reducing client 
    communication.
    Our evaluations on 300-replica geo-distributed deployment reveal that 
    \theparadigm protocols achieve significant performance gains over prior art: 
    $\approx$3x better throughput and $\approx$50\% better latency. 
\end{abstract}

\maketitle

\section{Introduction}
\label{sec:intro}

Byzantine consensus protocols are a perfect fit for solving the agreement 
problem in \emph{consortium} Blockchain platforms~\cite{androulaki2018hyperledger} 
due to their ability to shield the system from known but potentially 
mistrustful participants while reaching consensus efficiently, as opposed to 
Proof-of-Work-based~\cite{nakamoto2008bitcoin} techniques.
The fundamental requirement of any Blockchain platform is scalability to 
hundreds of nodes deployed around the world. 
Traditional Byzantine Fault-Tolerant (BFT) 
consensus protocols~\cite{castro1999practical,castro2002practical,cowling2006hq,kotla2007zyzzyva,kotla2009zyzzyva} 
suffer from intrinsic design issues that inhibit their scalability in 
geographically distributed (geo-distributed) deployments.

Most deterministic BFT consensus 
protocols~\cite{castro1999practical,cowling2006hq,kotla2007zyzzyva,Gutea:2019:SBFT} 
adopt the primary-backup approach, 
where a designated primary replica is responsible for ordering and replicating 
the client-submitted commands among the replicas. 
Relying on a dedicated replica to perform both these operations is detrimental to 
performance, especially at scale.
In particular, such an approach causes 
a) load imbalance among primary and backup replicas, 
because the primary sends larger messages containing client 
commands, while backups send small state messages;
b) under utilization of resources at backup replicas, 
because the primary saturates its network resources before the 
replicas, diminishing their individual potential;
c) remote clients to pay high WAN latencies by sending requests to the primary
than clients that are local to the primary;
and 
d) poor tolerance to primary failures~\cite{gupta2021rcc}.
Client commands in Blockchain applications (e.g. smart contracts) are 
typically large in the order of kilobytes
~\cite{androulaki2018hyperledger,bitcoin-tx-size}
limiting the number of commands that can be sent by primary using its bandwidth 
to \emph{all} replicas.

Existing BFT solutions that overcome these downsides of the primary-backup approach 
have drawbacks. Specifically, the rotating primary
~\cite{Veronese:2009:SOW:1637865.1638341,Veronese:2010:EEB:1909626.1909800,Yin:2019:HotStuff:3293611.3331591} 
and multi-primary~\cite{gupta2021rcc,Stathakopoulou:2019:MirBFT:arxiv:1906.05552} 
approaches
do not take into account many aspects of modern geographically 
distributed systems including variations in node hardware, network bandwidth, 
and available resources. 
In such settings, a slow node can quickly degrade the overall performance. 
Some decentralized approaches~\cite{ezbft,guerraoui2010next} exploit 
the commutativity of client commands and track dependencies 
to order conflicting commands. 
This requires additional coordination to process concurrent 
conflicting commands degrading performance.

\subsubsection*{Towards Partial Decentralization.}
To overcome these drawbacks, we present \theparadigm (for Divide and conQuer BFT), 
a paradigm for designing highly scalable consensus protocols by partially 
decentralizing the core consensus process 
into two distinct and concurrent steps that may be handled at potentially 
different replicas. Rather than adopting a completely decentralized approach 
where individual replicas replicate commands and also coordinate to find a 
\emph{total order}, 
\theparadigm divides the task of consensus into two: 
1) durable replication of client commands without a global order at correct replicas, 
and 
2) ordering of the commands to guarantee a \emph{total order}. 
Durable replication is carried out by each individual replica for the commands 
it receives from clients, while ordering is performed by a dedicated sequencer. 
Ordering involves assigning a \emph{global} order to a replica that has proposed 
a command. 
Thus, unlike the rotating primary and other multi-primary techniques
\cite{Veronese:2009:SOW:1637865.1638341,Yin:2019:HotStuff:3293611.3331591,gupta2021rcc,Stathakopoulou:2019:MirBFT:arxiv:1906.05552},
our approach can seamlessly accommodate variations in node hardware, network 
bandwidth, and available resources.

The \theparadigm approach is unique in that it allows for concurrent 
progression of the two stages
in the absence of failures. 
\theparadigm uses separate instances of consensus protocols at individual 
replicas to carry out replication  
providing load balancing of client commands among the replicas, 
while another consensus protocol is 
responsible for assigning the global order to individual replicas.  
This simultaneous replication and ordering allows 
\theparadigm to avoid the latency penalties due to the additional communication 
steps.
However, to limit the impact of Byzantine replicas in certain situations, 
\theparadigm requires that replication precede global ordering on a 
per-replica basis.
Decoupling replication from ordering has been proposed in the crash-fault 
model~\cite{Zhao:2018:SDPaxos:3267809.3267837,Li:2016:NoPaxos:OSDI:3026877.3026914,moraru2013there},
but, these protocols do not scale to hundreds of geo-distributed 
Byzantine replicas, require special network hardware, 
and/or are not oblivious to conflicts.

\subsubsection*{Towards Highly Scalable Consensus}

While the \theparadigm paradigm can be adopted into existing BFT protocols, 
we show, analytically in Figure~\ref{fig:analytical-model-plot} and 
empirically in Section~\ref{sec:eval}, that such instantiations do not 
scale their performance to hundreds of replicas.
Therefore, we present \thesystem, the flagship instantiation of the 
\theparadigm paradigm with three enhancements each of which contribute to 
achieve high performance while scaling to hundreds of replicas. 
Briefly, the techniques include: 
(1) using a hardware-assisted trusted subsystem to increase fault-tolerance 
and decrease quorum sizes; 
(2) linear communication for scalability; and 
(3) using threshold cryptography for optimal linear communication. 

BFT protocols require $3f+1$ replicas and three communication steps
among two-thirds of replicas to reach agreement.
In contrast, 
Hybrid consensus protocols~\cite{behl2017hybrids,veronese2013efficient} 
use trusted subsystems to require only $2f+1$ replicas and two communication 
steps among \emph{majority} replicas to reach agreement.
We show that such efficiency combined with the reduction in the number of 
messages exchanged per commit via linear communication patterns is key 
in leveraging the benefits of the \theparadigm paradigm at scale
(see Figures~\ref{fig:comparison-analysis-table} and 
\ref{fig:analytical-model-plot}).
With this insight, we adopt and linearize the common-case communication of 
a recent Hybrid protocol, Hybster~\cite{behl2017hybrids}, 
producing Linear Hybster. 
Both the replication and ordering steps of \thesystem use instances of 
\emph{Linear Hybster}. 

The ability of Hybrid protocols to tolerate more faults and use smaller size 
quorums enable scalability in geo-distributed environments.
Further, trusted execution environments are now available at
commodity-scale (e.g., Intel SGX~\cite{intel-sgx}, 
ARM's TrustZone~\cite{arm-trustzone}), making Hybrid protocols more feasible.
Regardless, the \theparadigm paradigm is generally applicable
to any BFT protocol and does not require Hybrid fault assumptions. 
\thesystem leverages \theparadigm and the Hybrid model to improve 
performance.
Many Blockchain solutions already depend on trusted execution environments for 
privacy-focused computations~\cite{10.1145/2976749.2978326, sgxwallet}, 
and thus, can easily take advantage of the added performance provided by Hybrid 
protocols.
       
\subsubsection*{Contributions} 
In Section~\ref{sec:bg}, we discuss the differences between BFT and 
Hybrid protocols and the challenges existing in the landscape.
In Section~\ref{sec:dqbft}, we propose \theparadigm, a paradigm for 
designing scalable BFT protocols 
by partially decentralizing the replication and ordering concerns. 
The technique can be applied to most primary-backup protocols 
to achieve high performance and scalability. 
In Section~\ref{sec:destiny}, we propose \thesystem, 
a Hybrid protocol under the \theparadigm paradigm that scales to 
hundreds of geo-distributed replicas.
In Section~\ref{sec:eval}, we present a comprehensive evaluation of 
the state-of-the-art protocols and four \theparadigm protocols, 
including \thesystem, in a geo-distributed deployment with various 
system sizes ranging from 19 up to 301 replicas, 
withstanding between $f=6$ and $150$ Byzantine failures. 
Our evaluations reveal that the \theparadigm variants of 
PBFT~\cite{castro1999practical}, SBFT~\cite{Gutea:2019:SBFT}, 
and Hybster~\cite{behl2017hybrids} -- 
DQPBFT, DQSBFT, and DQHybster -- outperform their 
vanilla counterparts with up to an order of magnitude better performance.
Furthermore, these protocols tolerate lagging replicas better than other 
multi-primary protocols with at least 20\% better throughput.
\thesystem provides $40\%$ better throughput than DQSBFT and 
up to $70\%$ lower latency than any other state-of-the-art protocol. 

\section{Background}
\label{sec:bg}

In this section, we provide the necessary background for understanding the rest
of the paper.

\subsection{Byzantine Consensus}

A Byzantine Fault-Tolerant (BFT) consensus protocol consists of a collection of 
replicas that agree on the order of client-issued commands and execute them in 
the agreed order. 
The protocol functions in a series of views, where in each view, a primary 
replica proposes and sequences commands, which are executed by all 
non-faulty replicas in the prescribed order.
Before executing the commands, correct replicas must ensure that 
(i) the commands are replicated at enough correct replicas and 
(ii) enough correct replicas observe the same sequence of commands 
from the primary. This function is carried out by the \emph{Agreement} 
algorithm, by exchanging command and state information between replicas. 
Some BFT agreement algorithms (e.g., PBFT~\cite{castro1999practical}) 
commit in three phases and require consent from a supermajority (i.e., 67\%) of replicas, while some others (e.g., SBFT~\cite{Gutea:2019:SBFT}) 
require consent from all replicas and commit in two phases during
non-faulty periods.

When the primary ceases to make timely progress or misbehaves by sending 
different sequence of commands to different replicas, 
the \emph{View Change} algorithm is invoked by non-faulty replicas to 
replace the faulty primary. 
The primary of the new view, determined by the view number, 
collects the replica-local states of enough replicas, 
computes the initial state of the new view, and 
proceeds with the agreement algorithm in new view. 
If a view change does not complete in time, another one is triggered 
for the next primary.

Replicas use the \emph{Checkpoint algorithm} to limit their memory 
requirements by garbage collecting the states for those commands that have been 
executed by enough correct replicas.
Replicas exchange information to produce the checkpoint state. 
When some replicas fall behind the rest of the system, the checkpoint state is 
used to bring them up to date via the \emph{state transfer} algorithm.

\subsection{Consensus with Trusted Subsystems}

Replicas in the BFT model may fail to send one or more messages specified by the 
protocol or even send messages not specified by the protocol.
These replicas can also equivocate, i.e., make conflicting statements 
to compromise consistency, without being detected.
To tolerate such behaviors, BFT protocols require at least supermajority 
\emph{quorums} — the subset of replicas that 
is used to make decisions at different phases of consensus — 
in asynchronous systems.

On the other hand, in the hybrid fault model, 
a trusted subsystem is employed to prevent replicas 
from equivocating~\cite{a2m, trinc}. 
A trusted subsystem is a local service that exists at every replica, and 
certifies the messages sent by the replicas to ensure that 
malicious replicas cannot cause different correct replicas to execute 
different sequences of operations. 
The trusted subsystem, typically, consists of a monotonically increasing counter 
that is paired with an attestation mechanism 
(signatures/message authentication codes).
The trusted subsystem assigns a unique counter to a message and generates a 
cryptographic attestation over the pair. 
Thus, each outbound message is bound to a unique counter value. 
When correct replicas receive the message pairs, they process them in 
increasing counter value order.
Thus, when a faulty replica sends two different messages to 
two different correct replicas, only one will process the message, 
while the other will wait for the message with the missing counter value, 
eventually detecting equivocation.

Since equivocation is prevented using the trusted subsystem, 
$f$ additional correct replicas that were required for 
traditional BFT protocols to balance the impact of $f$ malicious replicas 
are no longer required in the hybrid fault model. 
The result is smaller quorums. 
The system size ($N$) of 
traditional BFT protocols is $3f+1$; 
hybrid protocols improve this to $2f+1$.

\begin{figure}[t]
	\centering
	\begin{subfigure}{0.43\textwidth}
        \centering
		\includegraphics[width=\linewidth]{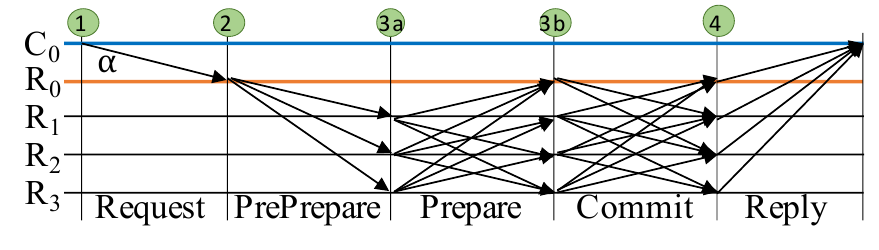}
		\vspace{-20pt}
		\caption{PBFT}
	\end{subfigure}
	\begin{subfigure}{0.33\textwidth}
        \centering
		\vspace{5pt}
		\includegraphics[width=\linewidth]{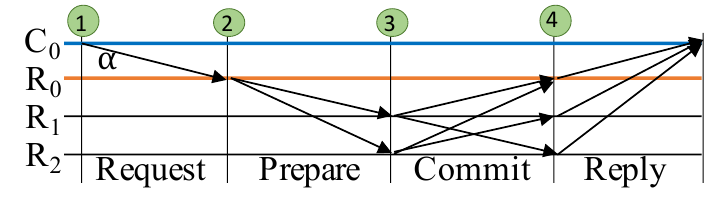}
		\vspace{-20pt}
		\caption{Hybster, hybrid protocol}
	\end{subfigure}
	\vspace{-5pt}
	\caption{Agreement protocol steps: PBFT and Hybster.}
	\label{fig:bg:walkthrough}
	\vspace{-5pt}
\end{figure}

\begin{example}
	\emph{BFT vs Hybrids.}
To illustrate the fundamental design differences between BFT and hybrid 
protocols, we compare PBFT (a BFT protocol) and Hybster~\cite{behl2017hybrids} 
(a hybrid protocol).
The agreement algorithm of PBFT and Hybster operates  
in three and two phases, respectively, as illustrated in 
Figure~\ref{fig:bg:walkthrough}.
Note that we describe the two protocols 
hand-in-hand and only highlight their differences. 

\circled{1} 
For each protocol, the execution starts when the client sends a command to 
the primary replica. The client signs its command to ensure that a malicious 
replica cannot tamper the command without detection. 
\circled{2} The primary receives the command and proposes it to all 
replicas with a sequence number by broadcasting a \texttt{Prepare} message. 
The sequence number defines the \emph{global} 
execution order with respect to other commands.
The message is certified using a \emph{message authentication code} (MAC). Hybster uses the trusted subsystem to produce the MAC. 
The counter value of Hybster maps to the sequence number assigned to the command. 
Thus, two different commands are never assigned the same sequence number in Hybster. 
Also, note that in the example, Hybster requires three nodes, while PBFT requires four.

The replicas receive the proposal from the leader. 
\circled{3} \underline{\textbf{Hybster}} replicas acknowledge the proposal to each other. Replicas wait for a majority of responses to commit and execute the command. 
\circled{3a}
\underline{\textbf{PBFT}} replicas exchange the proposal 
with each other to ensure that they received the same proposal 
from the primary (i.e., to ensure no equivocation). 
Note that Hybster avoids this step using the trusted subsystem.
The proposal is validated if a supermajority of nodes respond with the same proposal from the primary. 
\circled{3b} 
\underline{\textbf{PBFT}} replicas exchange commit messages. 
They execute the command upon collecting a majority quorum of these messages and reply to the client. 
At the end of this step, in both the protocols, a correct replica is able to 
recover the command, even if $f$ replicas fail including the primary.
\circled{4}
Clients wait until they receive identical replies from at least $f+1$ replicas. 
This is because, waiting for only one potentially malicious replica may 
yield an incorrect result.
\end{example}

\begin{figure*}[t]
	\begin{minipage}{.55\textwidth}
		\begin{tabular}{|cccc|}
			\hline
			Protocol & Messages & Throughput & Phases \\\hline
			PBFT & $N + 2N^2$ & $B/(N-1)(pm+3sm)$ & 3 \\\hline
			Hybster & $N + N^2$ & $B/(N-1)(pm+sm)$ & 2 \\\hline
			\makecell{DQPBFT\\(ours)} & $2N + 4N^2$ & \small \begin{tabular}{c}$NF*B$\\ \hline $(N-1)(pm+3sm)+(NF-1)$\\$(pm+4(N-1)sm) + (N-1)(4sm)$\end{tabular} & 4 or 6 \\\hline
			\makecell{Destiny\\(ours)} & $7N$ & \small \begin{tabular}{c}$NF*B$\\ \hline $(N-1)(pm+3sm)+(NF-1)$\\$(pm+3sm) + (N-1)(4sm)$\end{tabular} & 5 or 7 \\\hline
		\end{tabular}
		\vspace{-10pt}
		\caption{Comparison of single-primary and DQBFT protocols.\\
		$B$: bandwidth per replica; $N$: system size; $NF$: non-faulty replicas;\\ 
		$pm$: size of payload messages; $sm$: size of state messages.}
		\label{fig:comparison-analysis-table}
	\end{minipage}
	\begin{minipage}{.44\textwidth}
		\centering
		\includegraphics[width=0.7\textwidth]{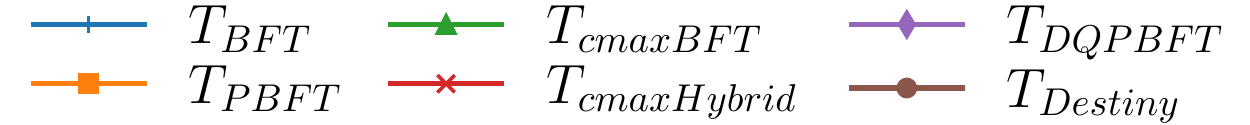}
		\begin{subfigure}[b]{0.5\columnwidth}
			\includegraphics[width=\textwidth]{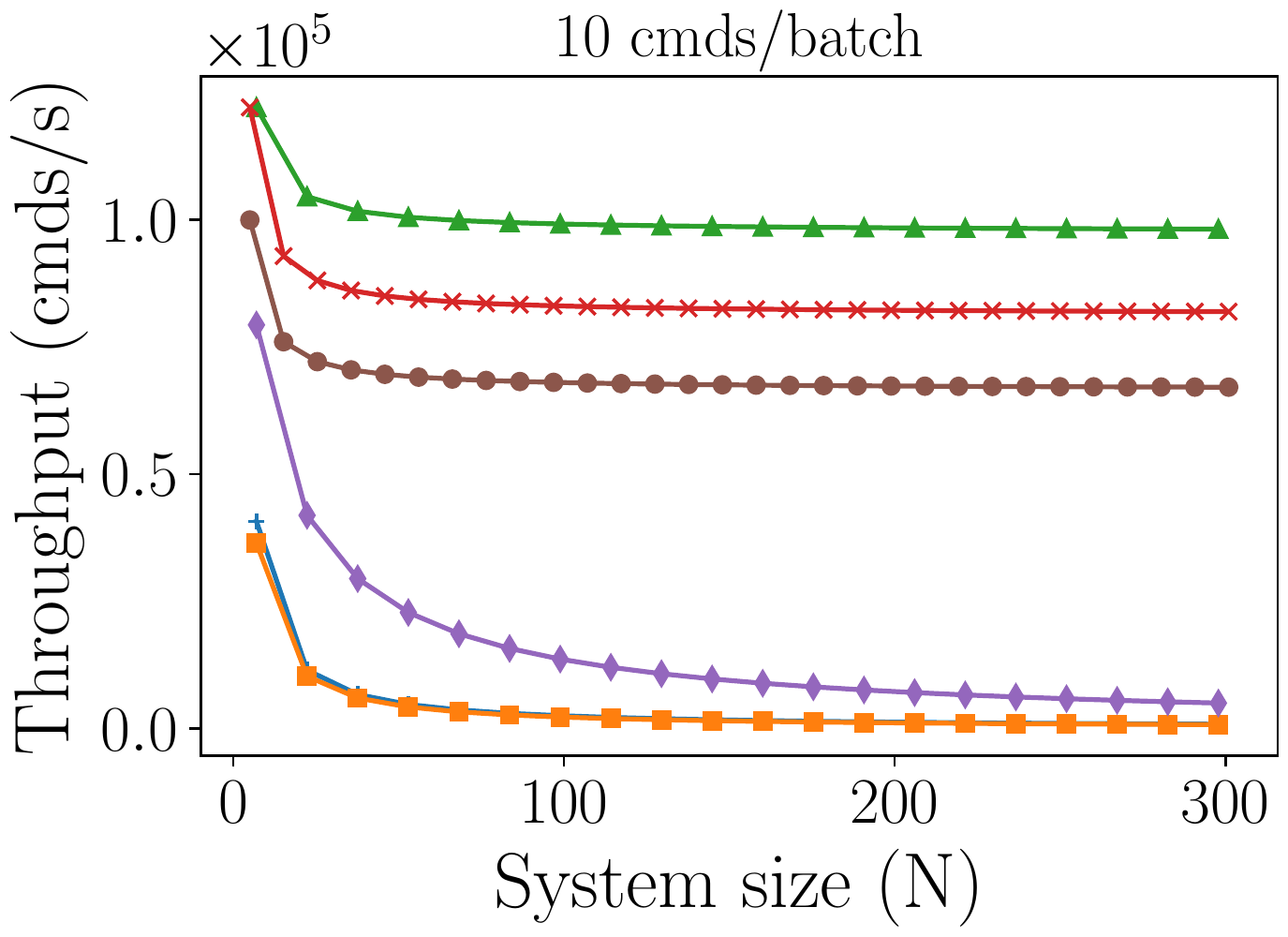}
		\end{subfigure}
		\begin{subfigure}[b]{0.49\columnwidth}
			\includegraphics[width=\textwidth]{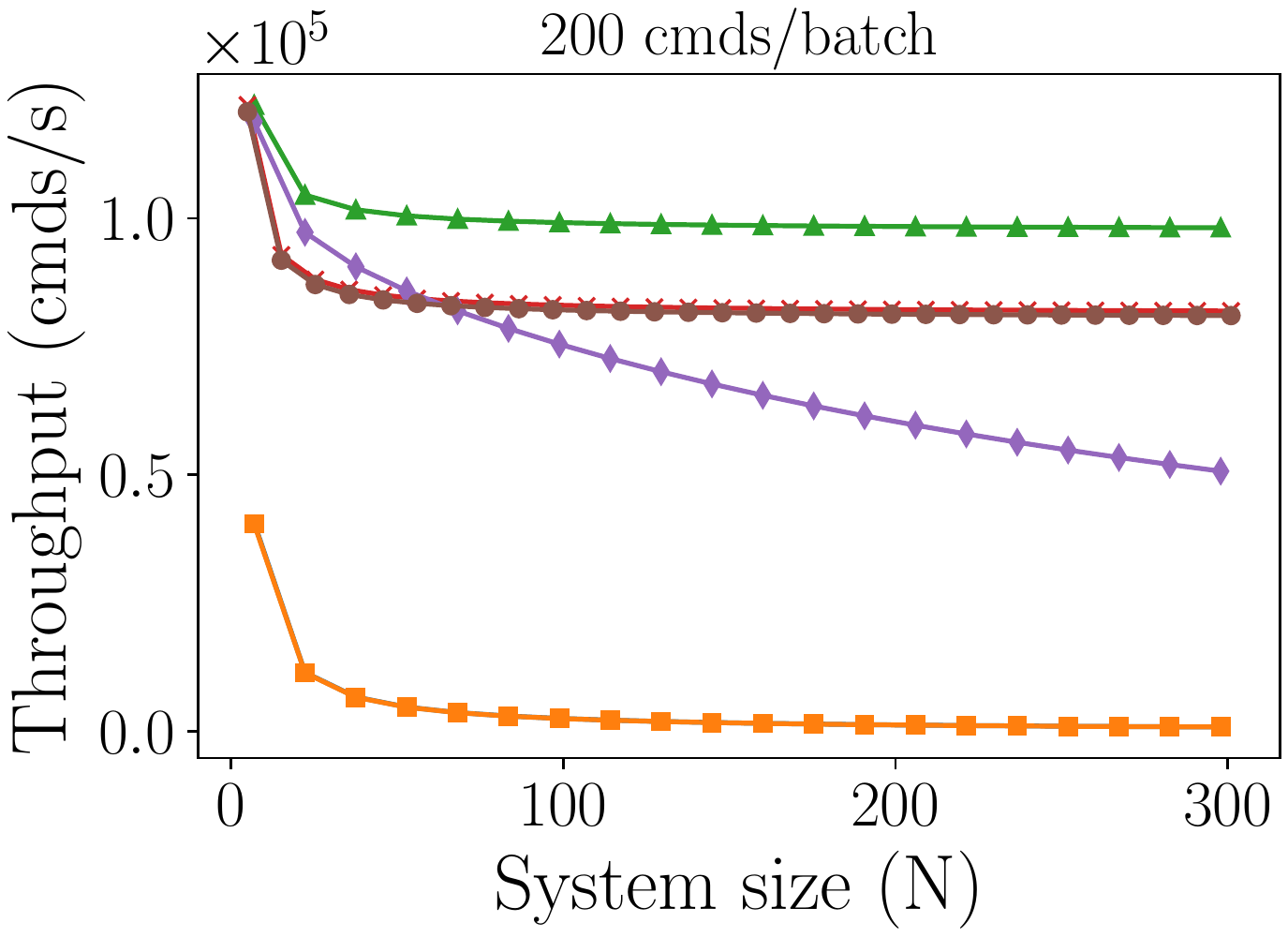}
		\end{subfigure}
		\vspace{-20pt}
		\caption{Maximum theoretical throughput in a system with $B=1Gbit/s$,
		$NF = N - f$, $sm = 250B$, and $pm = 5KiB$ (left) and $100KiB$ (right)
		respectively.}
		\label{fig:analytical-model-plot}
	\end{minipage}
\end{figure*}

\subsection{Decentralizing Consensus}
\label{sec:bg:dc}

A major problem with quorum-based BFT consensus protocols 
(e.g., PBFT~\cite{castro1999practical}, SBFT~\cite{Gutea:2019:SBFT}, 
Hotstuff~\cite{Yin:2019:HotStuff:3293611.3331591})
that underpin numerous  
Blockchain infrastructures~\cite{Bessani:2017:BFO:3152824.3152830} 
is their reliance on a designated replica, called the primary, 
to order client commands. 
The maximum theoretical throughput at which a primary can replicate client
commands is $T_{p} = B/((N-1)pm)$, where $B$ is the primary's network bandwidth 
and $pm$ is the size of payload message and $N$ is the number of replicas
~\cite{gupta2021rcc}.
However, to ensure safety, the primary/replicas exchange state messages 
with each other and these messages must be taken into account.
Figure~\ref{fig:comparison-analysis-table} presents the theoretical 
throughput equations for protocols including PBFT and Hybster and 
Figure~\ref{fig:analytical-model-plot} plots it for two payload sizes. 
Note that this throughput is based on replica bandwidth only; in practice, 
the throughput is also affected by available computation and memory resources. 
The primary sends $(N-1)pm$ bytes, while other replicas only receive 
roughly $pm$ bytes each, leading to load imbalance and underutilization of 
replica resources.
By distributing the primary's responsibility and allowing all replicas to 
replicate the command payloads concurrently, one can achieve maximum throughput 
$T_{cmax} = (NF{\cdot}B)/((N-1)pm+(NF-1)pm)$, 
where $NF$ is the minimum number of non-faulty replicas ($N-f$) and is 
different for BFT and Hybrid protocols.
The literature presents two main methodologies to accomplish this.

In the first approach, referred to as \emph{static} ordering, 
the sequence numbers used to order the commands  
are statically partitioned among replicas. 
Replicas use their allocated set of sequence numbers to propose and 
commit commands, either in parallel~\cite{gupta2021rcc} or 
in round-robin fashion~\cite{Yin:2019:HotStuff:3293611.3331591}.
To ensure linearizability~\cite{ahs+94}, 
replicas must execute commands in the order of their sequence numbers.
Such an approach cannot adapt to variations in node hardware 
and network bandwidth.
A slow replica can throttle the system performance as commands must 
be effectively executed at the speed at which the slowest replica can 
propose and commit commands.
Examples of protocols that adopt variants of this approach 
include Hotstuff~\cite{Yin:2019:HotStuff:3293611.3331591}, 
RCC~\cite{gupta2021rcc}, 
MirBFT~\cite{Stathakopoulou:2019:MirBFT:arxiv:1906.05552}, 
and Dispel~\cite{Voron:2019:Dispel:arxiv:1912.10367}.

In the second approach, referred to as \emph{dependency}-based ordering, 
replicas commit commands by exchanging dependency metadata,  
and execute those commands after deterministically ordering 
them using the agreed-upon dependency information. 
The order of execution of the commands depends on the nature of the operations 
in those commands. Commands with conflicting operations are totally ordered 
while others are partially ordered~\cite{Defago:2004:TOB:1041680.1041682}. 
Such dynamic ordering minimizes the overhead of ordering non-conflicting 
commands, because their reordering does not cause inconsistent system state. 
Such protocols~\cite{moraru-thesis,arun2019ezbft} incur higher overhead when 
the number of conflicting commands is high, degrading performance.
\section{The Divide and Conquer Paradigm}
\label{sec:dqbft}

We propose \theparadigm, a paradigm for building high-performance BFT protocols
that overcomes the aforementioned challenges in existing protocols.
To do so, \theparadigm decentralizes the responsibility of the primary 
based on the two important actions performed by a consensus protocol:
i) request dissemination with partial ordering and 
ii) global ordering. 
Request dissemination is a decentralized operation and  does not require 
replicas to coordinate, but only acknowledge receipt. 
In contrast, global ordering requires replicas to coordinate to ensure that 
the system has a single view of the sequence of operations. 
To simplify this process, a replica is elected to propose the global ordering 
for the commands.

Under the \theparadigm paradigm, clients can send commands to any replica. 
Replicas can individually disseminate and order client commands using multiple 
instances of a consensus protocol. 
While this ensures that the commands are disseminated to the correct replicas, 
the order produced is local to the replica, i.e. \emph{a partial order}, 
and not global. 
The \emph{primary} replica produces a global order among the partial orders 
produced by the individual replicas.

Such a \emph{partially-decentralized} approach has the following benefits. 
First, the decentralized process of dissemination distributes load evenly 
across replicas and enables clients to connect to the nearest replica in 
geo-distributed deployments. Second, for ordering, the global view of all commands enables the sequencer to order them optimally,  i.e., each newly proposed command can be dynamically assigned to the first unused global sequence number. Unlike other multi-primary~\cite{gupta2021rcc,Stathakopoulou:2019:MirBFT:arxiv:1906.05552} 
and rotating-leader~\cite{Yin:2019:HotStuff:3293611.3331591,Veronese:2010:EEB:1909626.1909800,Veronese:2009:SOW:1637865.1638341} protocols whose performances suffer due to slow replicas, our technique allows replicas to execute commands 
at their own pace without being bottle-necked by slower replicas. 
Moreover, such an approach is oblivious to conflicts (unlike, for 
e.g., \textsc{ezBFT}~\cite{ezbft}, Aliph~\cite{guerraoui2010next}). 

\subsection{Design}

At a high level, the \theparadigm paradigm is composed of two sub-protocols:  
the dissemination protocol and the global ordering protocol. The dissemination protocol employs multiple instances of consensus, called D-instances, to enable every replica to disseminate and partially order its client commands. Meanwhile, the global ordering protocol uses a single instance of consensus, called the O-instance, that agrees on the global order among the partial orders produced by the D-instances. There are as many D-instances as there are replicas, and every replica is the coordinator of at least one D-instance.
A replica proposes commands in a series of sequence numbers belonging 
to its own D-instance to produce its partial order. 
The primary proposes D-instance sequence numbers in O-instance's sequence number 
space to effectively produce a global order from the replica-specific partial orders.

To tolerate Byzantine faults, both dissemination and ordering should be based 
on a BFT consensus protocol. 
Primary-based protocols with the following properties~\cite{veronese2013efficient} 
can be used for instantiating protocols under the \theparadigm paradigm.
\begin{enumerate}[nosep, label=(P\arabic*), ref=P\arabic*]
    \item\label{prop:bft1} 
    If a correct replica executes a command $\alpha$ 
    at sequence number $\mathcal{S}$ 
    in view $v$, 
    no correct replica will execute $\alpha' \neq \alpha$ with sequence number 
    $\mathcal{S}$.
    \item\label{prop:bft2} 
    If a correct replica executed a command $\alpha$ 
    at sequence number $\mathcal{S}$ 
    in view $v$, 
    no correct replica will execute $\alpha$ with sequence number 
    $\mathcal{S}' > \mathcal{S}$
    in any view $v' > v$.
    \item\label{prop:bft3} 
    During a stable view where the communication between correct replicas is  
    synchronous, a proposed client command is committed by a correct replica.
    \item\label{prop:bft4}
    A view $v$ will eventually transition to a new view $v' > v$ if enough 
    correct replicas request for it.
\end{enumerate}

Many primary-based BFT and Hybrid protocols provide these properties 
and can be instantiated under the \theparadigm paradigm
~\cite{behl2017hybrids,castro2002practical,Gutea:2019:SBFT,kotla2007zyzzyva}. 
In Section~\ref{sec:eval}, we evaluate four such instantiations.

In \theparadigm, the O-instance primary replica performs a fixed amount of work. 
Since the task of disseminating the client commands is offloaded to other 
replicas, the primary is only responsible for total-ordering. 
Furthermore, the use of consensus protocols for both D-instances and 
the O-instances allows the dissemination and 
the global ordering steps to proceed simultaneously.
While dissemination is in progress, the ordering protocol 
\emph{optimistically} proposes a global ordering for the command. 
This allows for the communication steps of both protocols to overlap, 
and thereby effectively reduces the overall number of communication steps. 
Note that such concurrent processing does not strain the communication channels. 
Since the ordering protocol is only ordering the sequence numbers, 
the message sizes are constant and are only a few bytes. 
The dissemination protocol carries a larger and variable payload containing 
the client commands.
See Figure~\ref{fig:comparison-analysis-table} for a comparison of 
DQBFT protocols and Figure~\ref{fig:analytical-model-plot} for theoretical 
throughput analysis.

\subsection{\theparadigm}

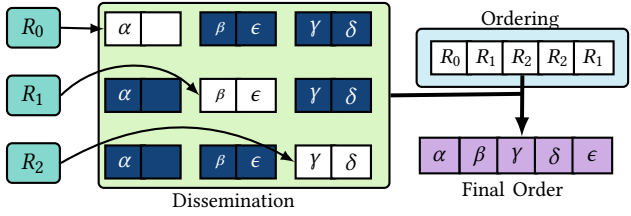
\begin{figure}[t]
    \centering
    \usetikzlibrary{arrows}
    \begin{tikzpicture}

        \definecolor{mygreen}{RGB}{132, 216, 205}
        \definecolor{mygreen2}{RGB}{219, 246, 197}
        \definecolor{myblue}{RGB}{19, 67, 121}
        \definecolor{skyblue}{RGB}{209, 238, 247}
        \definecolor{mypink}{RGB}{206, 173, 228}
        
        \node[text width=1.5cm] at (2.6,-0.4) {\small Dissemination};
        \node[text width=1.5cm] at (6.7,2) {\small Ordering};
        \node[text width=2cm] at (6.7,-0.25) {\small Final Order};
        
        \node[rectangle, draw=black, fill = mygreen, rounded corners = 2pt, thick, minimum height = 15pt, minimum width = 20pt] (r0) at (0,1.9) {$R_0$};
        \node[rectangle, draw=black, fill = mygreen, rounded corners = 2pt, thick, minimum height = 15pt, minimum width = 20pt] (r1) at (0,1) {$R_1$};
        \node[rectangle, draw=black, fill = mygreen, rounded corners = 2pt, thick, minimum height = 15pt, minimum width = 20pt] (r2) at (0,0.1) {$R_2$};
        
        \node[rectangle, draw=black, fill = mygreen2, rounded corners = 2pt, thick, minimum height = 70pt, minimum width = 110pt] (main) at (2.8,1) {};
        
        \node[rectangle, color = white, draw=black, fill = myblue, thick, minimum height = 12.5pt, minimum width = 15pt] at ([xshift = -1em, yshift = 2.8em]main.center) {\tiny $\beta$};
        \node[rectangle, color = white, draw=black, fill = myblue, thick, minimum height = 12.5pt, minimum width = 15pt] at ([xshift = 0.5em, yshift = 2.8em]main.center) {$\epsilon$};
        
        \node[rectangle, draw=black, fill = white, thick, minimum height = 12.5pt, minimum width = 15pt] (b) at ([xshift = -1em] main.center) {\tiny $\beta$};
        \node[rectangle, draw=black, fill = white, thick, minimum height = 12.5pt, minimum width = 15pt] at ([xshift = 0.5em]main.center) {$\epsilon$};
    
        \node[rectangle, color = white, draw=black, fill = myblue, thick, minimum height = 12.5pt, minimum width = 15pt] at ([xshift = -1em, yshift = -2.8em] main.center) {\tiny $\beta$};
        \node[rectangle, color = white, draw=black, fill = myblue, thick, minimum height = 12.5pt, minimum width = 15pt] at ([xshift = 0.5em, yshift = -2.8em]main.center) {$\epsilon$};
        
        \node[rectangle, draw=black, fill = white, thick, minimum height = 12.5pt, minimum width = 15pt] (a) at ([xshift = -5em, yshift = 2.8em]main.center) {$\alpha$};
        \node[rectangle, draw=black, fill = white, thick, minimum height = 12.5pt, minimum width = 15pt] at ([xshift = -3.5em, yshift = 2.8em]main.center) {};
        
        \node[rectangle, color = white, draw=black, fill = myblue, thick, minimum height = 12.5pt, minimum width = 15pt] at ([xshift = -5em]main.center) {$\alpha$};
        \node[rectangle, color = white, draw=black, fill = myblue, thick, minimum height = 12.5pt, minimum width = 15pt] at ([xshift = -3.5em]main.center) {};
        
        \node[rectangle, color = white, draw=black, fill = myblue, thick, minimum height = 12.5pt, minimum width = 15pt] at ([xshift = -5em, yshift = -2.8em]main.center) {$\alpha$};
        \node[rectangle, color = white, draw=black, fill = myblue, thick, minimum height = 12.5pt, minimum width = 15pt] at ([xshift = -3.5em, yshift = -2.8em]main.center) {};
        
        
        
        \node[rectangle, color = white, draw=black, fill = myblue, thick, minimum height = 12.5pt, minimum width = 15pt] at ([xshift = 3em, yshift = 2.8em]main.center) {$\gamma$};
        \node[rectangle, color = white, draw=black, fill = myblue, thick, minimum height = 12.5pt, minimum width = 15pt] at ([xshift = 4.5em, yshift = 2.8em]main.center) {$\delta$};
        
        \node[rectangle, color = white, draw=black, fill = myblue, thick, minimum height = 12.5pt, minimum width = 15pt] at ([xshift = 3em]main.center) {$\gamma$};
        \node[rectangle, color = white, draw=black, fill = myblue, thick, minimum height = 12.5pt, minimum width = 15pt] at ([xshift = 4.5em]main.center) {$\delta$};
        
        \node[rectangle, draw=black, fill = white, thick, minimum height = 12.5pt, minimum width = 15pt] (c) at ([xshift = 3em, yshift = -2.8em]main.center) {$\gamma$};
        \node[rectangle, draw=black, fill = white, thick, minimum height = 12.5pt, minimum width = 15pt] at ([xshift = 4.5em, yshift = -2.8em]main.center) {$\delta$};
        
       
       
       
        \node[rectangle, draw=black, fill = skyblue, rounded corners = 2pt, thick, minimum height = 22pt, minimum width = 80pt] (bluenode) at (6.5,1.5) {};
       
      \node[rectangle, draw=black, fill = white, thick, minimum height = 12.5pt, minimum width = 15pt] at ([xshift = -3em] bluenode.center) {\small $R_0$};
      \node[rectangle, draw=black, fill = white, thick, minimum height = 12.5pt, minimum width = 15pt] at ([xshift = -1.5em] bluenode.center) {\small $R_1$};
      \node[rectangle, draw=black, fill = white, thick, minimum height = 12.5pt, minimum width = 15pt] (main1) at ([xshift = 0em] bluenode.center) {\small $R_2$};
      \node[rectangle, draw=black, fill = white, thick, minimum height = 12.5pt, minimum width = 15pt] at ([xshift = 1.5em] bluenode.center) {\small $R_2$};
      \node[rectangle, draw=black, fill = white, thick, minimum height = 12.5pt, minimum width = 15pt] at ([xshift = 3em] bluenode.center) {\small $R_1$};
     
     \node[rectangle, draw=black, fill = mypink, thick, minimum height = 13.5pt, minimum width = 15pt] at (5.4,.2) {$\alpha$};
     \node[rectangle, draw=black, fill = mypink, thick, minimum height = 13.5pt, minimum width = 15pt] at (5.93,.2) {\small $\beta$};
     \node[rectangle, draw=black, fill = mypink, thick, minimum height = 13.5pt, minimum width = 15pt] (main2) at (6.43,.2) {$\gamma$};
     \node[rectangle, draw=black, fill = mypink, thick, minimum height = 13.5pt, minimum width = 15pt] at (6.93,.2) {$\delta$};
     \node[rectangle, draw=black, fill = mypink, thick, minimum height = 13.5pt, minimum width = 15pt] at (7.43,.2) {$\epsilon$};
     
     \path [->,>= latex, thick] (r0.east) edge node [left] {} (a.west);
     \path [->,>= latex, thick] (r1.east) edge[bend left = 45] node [left] {} (b.west);
     \path [->,>= latex, thick] (r2.east) edge[bend left = 30] node [left] {} (c.west);
     
    \draw [-, line width = 1.5] (main.east) to ([xshift = 0.2em, yshift=1.8em]main2.north);
    \draw [->, >=latex, line width = 1.5] (main1.south) to ([xshift = 0.2em]main2.north);
     
    \end{tikzpicture}
    \vspace{-5pt}
    \caption{\theparadigm's dissemination and ordering steps.}
    \label{fig:separating-concerns}
\end{figure}

In this section, we describe the inner workings of the \theparadigm paradigm 
and show how it accomplishes its goal of decentralizing the dissemination and 
global ordering steps, while preventing 
slow replicas from bottle-necking 
the system performance.
For the sake of exposition, we describe \theparadigm by applying it to PBFT. 
Figure~\ref{fig:separating-concerns} illustrates \theparadigm's separation of 
dissemination and ordering steps. 

\subsubsection{Agreement Protocol.}

\begin{figure*}
    \begin{mdframed}
    A DQPBFT replica executes the following sub-protocols:
    
    \textbf{Dissemination Protocol}
    
    $N$ instances of the PBFT protocol are used for dissemination. 
    Each replica ``owns'' one instance and replicates its client commands 
    with that instance. 
    The prefix "D-" and the replica identifier embedded in the messages helps 
    to identify the protocol instance.

    \begin{enumerate}[label=(\arabic*), leftmargin=*]
        \item \textbf{D-PrePrepare.} A replica $R_n$ receives a client command 
        $\alpha$ and sends a 
        $\langle \texttt{D-PrePrepare}, v_n, R_n, \mathcal{S}_{ni}, \alpha \rangle$ 
        message to all replicas. 
        $v_n$ is the view number and 
        $\mathcal{S}_{ni}$ is the lowest available sequence number.
        \item \textbf{D-Prepare.} 
        A replica $R_m$ that receives a PrePrepare message 
        $\langle \texttt{D-PrePrepare}, v_n, R_n, \mathcal{S}_{ni}, \alpha \rangle$ 
        ensures the validity of the view and sequence numbers.
        Consequently, $R_m$ sends a 
        $\langle \texttt{D-Prepare}, v_n, R_n, \mathcal{S}_{ni}, Hash(\alpha) \rangle$ 
        message to all the replicas. 
        \item \textbf{D-Commit.} A replica that collects $2f+1$ valid
        \texttt{D-Prepare} messages, sends the 
        $\langle \texttt{D-Commit}, v_n, R_n, \mathcal{S}_{ni}, Hash(\alpha) \rangle$ 
        message.
        A replica that receives $2f+1$ valid Commit messages marks the 
        operation as \emph{disseminated}.
    \end{enumerate}

    \textbf{Global Ordering Protocol}
    \begin{enumerate}[label=(\arabic*), leftmargin=*]
        \item \textbf{O-PrePrepare.} Case (i): If $R_n$ is in \emph{optimistic} 
        mode, then the primary $R_p$ of the global ordering protocol 
        assigns a global ordering number $\mathcal{S}_{pk}$ as soon as it
        receives the 
        $\langle \texttt{D-PrePrepare}, v_n, R_n, \mathcal{S}_{ni}, \alpha \rangle$  
        message from $R_n$.
        Case (ii): If $R_n$ is in \emph{pessimistic} mode, then $R_p$ assigns 
        $\mathcal{S}_{pk}$ only after the operation corresponding to 
        sequence number $\mathcal{S}_{ni}$ is marked as \emph{disseminated}.
        Once assigned, Primary $R_p$ sends the 
        $\langle \texttt{O-PrePrepare}, v_p, R_p, \mathcal{S}_{pk}, R_n, \mathcal{S}_{ni} \rangle$ 
        message to all replicas. 
        \item \textbf{O-Prepare.} A replica $R_q$ that receives the
        $\langle \texttt{O-PrePrepare}, v_p, R_p, \mathcal{S}_{pk}, R_n, \mathcal{S}_{ni} \rangle$ 
        message ensures the validity of the view and sequence numbers. 
        $R_q$ also ensures that there exists a corresponding 
        \texttt{D-PrePrepare} message, waiting if necessary. 
        It then sends a 
        $\langle \texttt{O-Prepare}, v_p, R_p, \mathcal{S}_{pk} \rangle$ 
        message to all the replicas. 
        \item \textbf{O-Commit.} A replica collects $2f+1$ valid
        \texttt{O-Prepare} messages, 
        and sends the
        $\langle \texttt{O-Commit}, v_p, R_p, \mathcal{S}_{pk} \rangle$ message.
        A replica that receives $2f+1$ valid Commit messages commits its 
        sequence number $\mathcal{S}_{pk}$ to map to $R_n$'s sequence number 
        $\mathcal{S}_{ni}$ and starts the execution procedure.
    \end{enumerate}

    \end{mdframed}
    \vspace{-5pt}
    \caption{\theparadigm Execution using the PBFT Protocol for both the
    D-instance and O-instance protocols.}
    \label{fig:dqbft-algorithm}
    \vspace{-5pt}
\end{figure*}

Figure~\ref{fig:dqbft-algorithm} presents the agreement protocol. 
We assume that a primary for the O-instance is elected beforehand. 
At a high level, a replica $R_n$ that receives the client command, say $\alpha$,
becomes the command's initial \emph{coordinator}. 
We say initial, because when the coordinator fails, it will be replaced using 
the View Change procedure.
The coordinator is responsible for partial ordering $\alpha$ 
with respect to the commands previously coordinated by it. 
The coordinator uses its D-instance protocol and assigns a sequence number 
$\mathcal{S}_{ni}$ and runs the consensus protocol to disseminate the command 
to other replicas.
Concurrently, the O-instance primary optimistically globally orders the 
D-instance sequence number $\mathcal{S}_{ni}$. 
The O-instance primary uses the PrePrepare message sent by the D-instance primary 
as the request for finding the global order for $\mathcal{S}_{ni}$.
Note that the O-instance protocol only orders the D-instance sequence numbers, 
only and not the commands. 
The O-instance primary $R_p$ assigns a sequence number $\mathcal{S}_{pk}$ to 
the D-instance number $\mathcal{S}_{pk}$ and 
sends protocol messages to replicas to produce a global order for $\alpha$.

\subsubsection{Execution.} 

A command $\alpha$ proposed by replica $R_n$ 
is \emph{decided} at a replica when it has been committed under a 
D-instance sequence number $\mathcal{S}_{ni}$, and 
$\mathcal{S}_{ni}$ has been 
committed under a O-instance sequence number $\mathcal{S}_{pk}$. 
However, the command cannot be executed yet. 
The command is considered \emph{ready} for execution only after all the 
corresponding commands mapped to the O-instance sequence numbers 
up to $\mathcal{S}_{pk}$ have been committed and executed.
Replicas execute the command and respond to the client.

\subsubsection{Checkpoint and State-transfer Protocols.}

As described previously, consensus protocols use the checkpoint mechanism to 
reduce the memory footprint at the replicas by garbage collecting logs 
for previously executed commands. This procedure also aids in bringing any lagging replicas to the latest state, by allowing up-to-date replicas to exchange the checkpoint data and recent logs via the state-transfer protocol. Note that replicas can be lagging as a result of the primary's intentions. This has implications for \theparadigm.

\begin{example}
    Consider a \theparadigm instantiation of PBFT with
    $N=4$ ($f=1$) replicas. 
    Let $R_0$ be the O-instance primary and be Byzantine. 
    Let $R_1$ and $R_2$ use their D-instances to commit two commands, 
    say, $\alpha$ and $\beta$, by sending them only to 
    quorum replicas $R_0$, $R_1$, and $R_2$.
    Let $R_0$, the O-instance primary, order the D-instances 
    using quorum replicas $R_0$, $R_1$, and $R_3$. 
    Now, $R_1$ is the only correct replica that can execute both the commands 
    and respond to the client. 
    Neither $R_2$ nor $R_3$ have all the necessary O-instance and D-instance 
    messages, respectively, to execute the commands.
    Thus, these replicas must transfer up-to-date state from other correct 
    replicas before execution. 
    \label{ex:byz-msg-witholding}
\end{example}

We now discuss measures to prevent malicious replicas from stalling the 
progress of other replicas.

\subsubsection{Controlling Byzantine Behavior.}
\label{sec:dqbft:byz-behavior}

Although the optimistic ordering mechanism reduces the number of effective 
communication steps, Byzantine replicas can still prolong the latency 
by refusing to send messages (see Example~\ref{ex:byz-msg-witholding}), 
thereby negatively affecting performance. 
A common technique to prevent this behavior is by flooding the 
\texttt{D-PrePrepare} messages~\cite{Amir:2011:Prime:TDSC.2010.70}. 
When a correct replica receives a \texttt{D-PrePrepare} message, 
it will multicast the message to all other replicas. 
This will ensure that the \texttt{D-PrePrepare} message will be received 
by other replicas in one communication step after it is initially
received by a correct replica. 
In larger systems, we observed that using a random subset of few replicas 
was equally effective while reducing the additional bandwidth requirements.

Despite such techniques, a coordinator can still collude with the primary and 
cause a global ordering slot to be committed without disseminating the command. 
This will eventually cause a view change (described below), which when 
frequent can reduce the overall performance of the system.
To prevent this behavior, we fall back to a more pessimistic approach on a per 
D-instance basis. 

After a view change, the D-instance will be placed under \emph{probation}, 
during which the O-instance primary will assign sequence numbers only 
pessimistically after its commands are disseminated. 
If the D-instance appears to behave for a certain period, the optimistic mode 
is restored. 
The period is denoted using sequence numbers that 
exponentially increases with each view change.
If correct replicas identify that the O-instance primary assigns 
sequence numbers optimistically for a D-instance on probation period even 
after some grace period, the O-instance primary will be replaced via a 
view change. 
During this time, correct replicas will only respond after the respective 
command becomes disseminated.

\subsubsection{View Change Protocol}
\label{sec:dqbft:vc}

There are cases where the coordinator or the primary does not make progress, 
either deliberately or due to other non-Byzantine causes 
(e.g., network disruptions). 
The view change protocol is used to reinstate progress whenever D-instance
coordinator or the O-instance primary fails to do so. 
One of the important characteristics of \theparadigm is that it adds no 
additional complexity to the existing view change procedure of the 
underlying consensus protocols. 
We assume an eventually synchronous~\cite{cachin2011introduction} 
communication between replicas where messages can be lost, be arbitrarily 
delayed, or arrive in any order, so
it is impossible to distinguish a Byzantine primary or coordinator 
that does not send any messages from a network fault. 
Thus, such protocols can only provide progress in periods during which the 
system is synchronous and messages arrive in bounded time.

Since, in \theparadigm, replicas can serve multiple roles (e.g., primary, 
coordinator, backup) at the same time, it is possible that a replica makes 
progress on a subset of roles while ceasing progress on other roles. 
By using the view change procedures of the respective D-instances and 
the O-instance, we ensure that only those primary and coordinator roles that 
do not make progress are replaced, without affecting the other roles. 
The view change can cause a replica to coordinate multiple D-instances 
including its own, however a replica is allowed to propose new commands 
using only its D-instance.

\textbf{Case 1: D-instance fails but O-instance is active.} 
A client sends its command to its \emph{assigned} coordinator.
If it does not receive $f+1$ responses for its command in time, 
it forwards the command to all replicas periodically. 
If timeouts happen often, a correct client can adapt by sending future commands 
to $f$ or more replicas.
Replicas will respond to the client if they have a reply.
Correct replicas that have not yet seen the D-instance sequence number assigned 
for the command will forward the command to the target coordinator 
and wait for the coordinator to assign a sequence number under its D-instance 
and send the initial message. If the timers expire before receiving the 
message, correct replicas will invoke the view-change procedure for 
that D-instance. 
The failure of a D-instance does not affect the O-instance progress, 
but can affect the execution phase. 
With the optimistic mode, it is possible that the O-instance primary globally 
orders the D-instance sequence number, but the sequence number 
did not commit before the view change, and no correct replica is aware of the 
command in that sequence number.
Thus, total ordering of command and execution must wait until a new 
coordinator is chosen for the D-instance, and it disseminates either a command 
or a special \emph{no-op} command.
The \emph{no-op} is proposed for all for sequence numbers that do not have 
a command associated but were committed in the O-instance

\textbf{Case 2: O-instance fails but D-instances are active. }
When the O-instance primary fails, D-instances will continue disseminating 
commands, but they will not be globally ordered.
After the O-instance undergoes a view change, D-instances must send their 
request to the new O-instance primary.
The respective D-instance coordinator and a subset of correct replicas 
(see Section~\ref{sec:dqbft:byz-behavior})
will periodically send the D-PrePrepare to the new primary until the 
O-PrePrepare is received. 
A client can also time out if the O-instance primary fails 
to make timely progress. 
Correct replicas monitor the O-instance primary to ensure that it 
assigns corresponding global sequence numbers for those commands that have been 
\emph{disseminated} in time. 
A view-change is triggered if the timer expires.

\textbf{Case 3: Both O-instance and D-instance fail simultaneously.}
The failure of the O-instance primary or a D-instance coordinator does not 
affect other active D-instances from disseminating commands. 
We run the view change protocols for the failed instances individually.
If the D-instance finishes view change before O-instance does, 
it can continue disseminating new commands (same as Case 2).
If the O-instance finishes view change before D-instance does, 
the O-instance will receive and order the sequence numbers for active 
D-instances (same as Case 1).
If the new primary or coordinator fails to make progress, the respective 
instance undergoes another view change.

When a previously failed replica restarts, the view-change protocol is used to 
reinstate the replica's D-instance, i.e. make the original replica the 
coordinator of its D-instance. 
After the replica restarts, other replicas will trigger a view change, 
skipping views if necessary, to reinstate the replica immediately. 
Note that $f+1$ replicas must agree to skipping views, so Byzantine replicas 
alone cannot reinstate. 
A correct replica will ensure that a recovered replica is participating in 
the protocol as a health check before agreeing to the view-change. 
Once reinstated, the replica must face probation.

\subsubsection{Client.}
A client command contains an operation and a monotonically increasing timestamp. 
Every replica caches the last executed timestamp and the reply 
for each client.
This is used to ensure that the replicas do not execute duplicate operations 
and to provide a reply to the client when required.
Similar to other multi-primary protocols, 
each client is assigned to a replica to prevent request duplication attacks, 
where faulty clients can send duplicate commands to multiple replicas 
simultaneously.
Even though replicas deduplicate commands during execution preserving safety,
it can nullify the throughput improvements achieved by using multiple 
primaries.
In \theparadigm, this assignment is carried out by running consensus 
on a special \texttt{ASSIGN} message via the O-instance.

\subsection{Correctness}

\theparadigm guarantees the following properties of a consensus protocol:
\begin{compactitem}[-]
    \item \textbf{Safety.} Any two correct replicas will execute the same sequence of client requests.
    \item \textbf{Liveness.} A client request proposed by a replica will eventually 
    be executed by every correct replica.
\end{compactitem}

\begin{lemma}\label{lm:destiny:s1}
    If a correct replica executes a command $\alpha$ whose D-instance 
    sequence number $\mathcal{S}_{ni}$ is mapped to O-instance sequence number 
    $\mathcal{S}_{pk}$ in view $v$, no correct replica will execute 
    $\beta \neq \alpha$ at O-instance sequence number $\mathcal{S}_{pk}$ in view $v$.
\end{lemma}
\begin{proof}
    The D-instance and O-instance protocols satisfy 
    Property \ref{prop:bft1} (Section~\ref{sec:dqbft}).
    Consequently, $\alpha$ is committed by replica $R_n$'s D-instance at 
    sequence number $\mathcal{S}_{ni}$ by correct replicas.
    Furthermore, $\mathcal{S}_{ni}$ is the value committed at O-instance number 
    $\mathcal{S}_{pk}$.
    Assume that a correct replica $R_m$ executes $\beta$ at $\mathcal{S}_{pk}$. 
    This would entail that either (i) $\beta$ was committed at $\mathcal{S}_{ni}$
    by correct replicas, or (ii) some $\mathcal{S}_{nj}$ assigned to $\beta$ was 
    committed at O-instance $\mathcal{S}_{pk}$ instead of $\mathcal{S}_{ni}$. 
    This contradicts Property \ref{prop:bft1}.
\end{proof}

\begin{lemma}\label{lm:destiny:s2}
    If a correct replica executes a command $\alpha$ whose D-instance 
    sequence number $\mathcal{S}_{ni}$ is mapped to O-instance sequence number 
    $\mathcal{S}_{pk}$ in view $v$, no correct replica will execute 
    $\beta \neq \alpha$ at O-instance sequence number $\mathcal{S}_{pk}$ 
    in any view $v' > v$.
\end{lemma}
\begin{proof}
    The individual consensus instances satisfy Property \ref{prop:bft2}. 
    The property ensures that if a command is chosen at a sequence number, 
    it will remain chosen at that sequence number at all higher views. 
    Thus, $\alpha$ remains chosen at $\mathcal{S}_{ni}$ at all higher views. 
    Similarly, $\mathcal{S}_{ni}$ is mapped to $\mathcal{S}_{pk}$ at all higher 
    views. 
    Suppose a correct replica executes $\beta$ at view $v' > v$. Then either 
    (i) $\beta$ is assigned to $\mathcal{S}_{ni}$ by correct replicas, or 
    (ii) some $\mathcal{S}_{nj}$ assigned to $beta$ was committed at 
    O-instance $\mathcal{S}_{pk}$ instead of $\mathcal{S}_{ni}$. 
    Both the conditions contradict Property \ref{prop:bft2}.
\end{proof}

Lemmas~\ref{lm:destiny:s1} and~\ref{lm:destiny:s2} satisfy the following 
theorem for safety:
\begin{theorem}\label{th:duobft:s1}
    Any two correct replicas commit the same sequence of operations. 
\end{theorem}
\begin{lemma}\label{lm:destiny:l1}
    During a stable view of the O-instance and the D-instance, 
    a proposed client command is executed by a correct replica.
\end{lemma}
\begin{proof}
    In a stable view, the correct primary will propose client requests 
    in a timely fashion to the replicas (Property~\ref{prop:bft3}). 
    Thus, the D-instance primary will ensure dissemination of the client 
    requests.
    Since there are at most $f$ faulty replicas, there will remain $N - f$ 
    correct ones that will respond to the  primary's messages.
    Thus, the client commands will be committed during the view 
    by correct replicas after receiving from a correct D-instance primary.
    The O-instance primary, being one of the correct replicas, will receive 
    the D-instance primary's \textsc{PrePrepare}
    with the client command and sequence number. 
    The correct O-instance primary will send the D-instance sequence number as 
    its command to correct replicas, and it will be committed in the view. 
    Thus, the client command will be assigned a global order by correct 
    replicas mapping the D-instance sequence number to the corresponding 
    O-instance.
\end{proof}

\begin{lemma}\label{lm:destiny:l2}
    A view $v$ will eventually transition to a new view $v' > v$ 
    if at least $N-f$ replicas request for it.
\end{lemma}
\begin{proof} 
    The proof follows directly from Property~\ref{prop:bft4} applied to 
    both the D-instances and the O-instance.
\end{proof}

\begin{theorem}
    A command sent by a correct client is eventually executed by 
    correct replicas.
\end{theorem}
\begin{proof}
    During a stable view, 
    Lemma~\ref{lm:destiny:l1} shows that the proposed command is 
    learned by the correct replicas. 
    When the view is unstable and the replica timers 
    expire properly, $f+1$ correct replicas will request a view change. 
    By Lemma~\ref{lm:destiny:l2}, a new view $v'$ will be installed.
    However, if less than $f+1$ replicas request the view change, then the 
    remaining replicas that do not request the view change will follow the 
    protocol properly. 
    Thus, the system will stay in view $v$ and the replicas
    will continue to commit commands in the view. 
    When proposals are not committed in time or 
    when more than $f$ replicas request a view change, 
    then all correct replicas will request a view change, and it will be 
    processed as in Lemma~\ref{lm:destiny:l1}.

    Even after a view change, the new view $v'$ may not necessarily be stable. 
    If the new primary deviates from the algorithm or does not make timely 
    progress, 
    correct replicas will request another view change 
    and move to the next view. 
    Since there can only be at most $f$ faulty 
    replicas, after at most $f+1$ view changes, 
    a stable view will be installed.
    Furthermore, if the faulty primary follows the algorithm enough such that 
    a view change cannot be triggered, by Lemma~\ref{lm:destiny:l1}, 
    replicas will continue to commit the commands.
\end{proof}

The individual consensus protocols satisfy 
linearizability~\cite{castro2002practical}.
The following theorem states that a command executed after committing via a 
D-instance and an O-instance satisfy linearizability.

\begin{theorem}
    \emph{Linearizability}: If $\alpha$ and $\beta$ are commands, 
    and the request for $\beta$ arrives after $\alpha$ is ready, 
    then $\alpha$ will be executed before $\beta$.
\end{theorem}
\begin{proof}
When $\alpha$ is ready, there must be at least $i$ O-instance sequence 
numbers belonging to $R_n$.
We prove this by contradiction.
Assume there are less than $i$ sequence numbers for $R_n$, but $\alpha$ is 
ready. 
This can happen only because there is a view change, and correct replicas 
observe less than $i$ sequence numbers. 
However, since $\alpha$ was ready for execution before the view change, 
there is at least one correct replica that will ensure that 
the primary of the new view enforces no less than $i$ instances, 
which is a contradiction.

When $\beta$ is received after $\alpha$ is ready, there should be at least $i$ 
O-instance sequence numbers committed belonging to $R_n$. There exists two cases. 
Case (i): If the O-instance primary is non-faulty, it will only assign 
sequence numbers in monotonically increasing order, 
so there will be no empty slots.
Case (ii): After a O-instance view change, correct replicas will observe 
at least $i$ sequence numbers belonging to $R_n$ since $\alpha$ is ready, and 
they will ensure that the new primary enforces the $i$ 
sequence numbers for $R_n$.
\end{proof}

\section{\thesystem}
\label{sec:destiny}

Although \theparadigm is a general paradigm that can benefit any primary-based BFT protocol, our performance evaluation (in Section~\ref{sec:eval}) reveals 
that not all protocols equally benefit from this approach. 
In this section, we present \thesystem, an instantiation of 
\theparadigm that is custom designed for scaling to hundreds of replicas, and achieve consistently high throughput and low latency even under high loads. \thesystem is able to take advantage of \theparadigm and achieve higher performance than state-of-the-art 
techniques~\cite{gupta2021rcc,Stathakopoulou:2019:MirBFT:arxiv:1906.05552}
at the scale of tens to hundreds of replicas.

\thesystem assumes the Hybrid fault model in order to tolerate more faults 
than BFT protocols for the same system size and also 
benefit from smaller quorums ($f+1$ instead of $2f+1$). 
\thesystem leverages a custom variant of Hybster~\cite{behl2017hybrids}, 
called Linear Hybster, to achieve its goal of higher performance and greater scalability.
Linear Hybster improves Hybster's normal-case communication complexity 
from quadratic to linear using threshold signatures and specialized collector 
roles. The collector aggregates messages from replicas 
and re-broadcasts them to all replicas. 
Since the messages are cryptographically signed, 
threshold signatures~\cite{Cachin:2005:ROC:2724955.2725009,Shoup:2000:PTS:1756169.1756190,stathakopoulous2017threshold} 
are used to reduce the number of outgoing collector messages from linear to 
constant. 
The same mechanism is employed for responding to the client. 
Clients wait for a single aggregated reply from a collector replica, 
instead of waiting for replies from $f+1$ replicas. 
The collector replica collects the signatures from $f+1$ replicas and sends a 
single response and signature to the client.

\subsection{Fault Assumptions and Cryptography}
\label{sec:model}

\thesystem assumes the Hybrid fault model -- 
the BFT model augmented with trusted hardware -- 
in which replicas can behave arbitrarily, 
except the trusted subsystem, which can only fail by crashing. 
Every replica, however, is capable of producing cryptographic 
signatures~\cite{Johnson:2001:ECD:2701775.2701951} that faulty replicas 
cannot break. 
We also assume a computationally bounded adversary 
that cannot do better than known attacks. 
The communication between replicas and clients is authenticated 
using public key infrastructures (PKI) such as TLS.
Being a hybrid protocol, \thesystem only requires $N = 2f + 1$ replicas 
to tolerate $f$ arbitrary failures. 

We consider an adversary that controls all the system software including the 
operating system. 
However, the adversary cannot read or modify the 
trusted subsystem's memory at run-time or 
decipher the secrets held inside it. 
Furthermore, the trusted subsystem is capable of generating cryptographic 
operations that the adversary cannot break. 
We also assume that the adversary cannot compromise the trusted subsystem's 
protections on participating nodes (e.g., via physical attacks). 
Preventing rollback attacks require replicating the subsystem state
~\cite{rote2017matetic}, which hybrid protocols perform during agreement 
implicitly.
Any compromise of the trusted component leads to safety violation 
of the protocol. 
Addressing this limitation is left as future work.

\thesystem uses threshold signatures to aggregate signatures at the collector. 
The threshold signature with a threshold parameter $t$ allows 
any subset $t$ from a total of $n$ signers to produce a valid signature 
on any message. 
It also ensures that no subset less than size $t$ can produce a valid signature. 
For this purpose, each signer holds a distinct private signing key that 
can be used to generate the corresponding signature share. 
The signature shares of a signed message can be combined into a single 
signature that can be verified using a single public key.
We use a threshold signature scheme based on Boneh-Lynn-Shacham (BLS) 
signatures~\cite{lynn2007implementation}. 
We use the BLS12-381~\cite{Barreto:2002:CEC:1766811.1766837} signature scheme 
that produces 192-byte signature shares. 
The aggregate signatures are also 192 bytes long. 

\subsection{The \thesubsystem Subsystem}

The \thesubsystem subsystem is a local service that exists on every replica. 
It allows for creating and verifying different types of threshold signatures 
for a message $m$ using a specified counter $tc$ and 
a corresponding counter value $tv$. 
By hosting part of \thesubsystem in a trusted subsystem, 
\thesubsystem guarantees a set of properties (described later) 
even if the replica is malicious.

\thesubsystem provides the following functions:
\begin{compactitem}[$\bullet$]
    \item \textbf{Independent Counter Signature Shares} 
    with input ($m$, $tc$, $tv'$). 
    \thesubsystem generates such a signature for a message $m$ if the provided 
    new value $tv'$ for counter $tc$ is \emph{greater than} 
    its current value $tv$. 
    It updates the counter $tc$'s value to $tv'$ and computes a signature share 
    using the subsystem's instance ID, counter $tc$'s ID, its new value $tv'$, 
    current value $tv$, and the message $m$.

    \item \textbf{Aggregate Signature Shares}. 
    It returns a single signature by aggregating at least $t$ 
    valid threshold signatures. 

    \item \textbf{Verify Signature}. It verifies the aggregated signature $sig$ 
    using the public key and indicates whether message $m$ was signed by 
    $t$ replicas with counter value $tv$ of counter $tc$.
    
    \item \textbf{Continuing Counter Certificates} with input 
    ($m$, $tc$, $tv$, $tv'$). \thesubsystem generates 
    a message authentication code (MAC) certificate for 
    a message $m$ if the submitted new value $tv'$ for 
    counter $tc$ is \emph{greater than or equal to} its current value $tv$. 
    It updates the counter $tc$'s value to $tv'$ and 
    computes a signature share using its private key share, 
    the subsystem's instance ID, counter $tc$'s ID, its new value $tv'$, 
    current value $tv$, and the message $m$. 
    
    \item \textbf{Verify Certificate} with input ($m$, $mac$, $tc$, $tv$, $tv'$). 
    It verifies the MAC certificate $mac$ using the secret key and returns true 
    if message $m$ is assigned a continuing certificate that transitions the 
    counter $tc$ from $tv$ to $tv'$.
\end{compactitem}

\thesubsystem also provides the capabilities of 
TrInX~\cite{trinc,behl2017hybrids}, 
the original Hybster's trusted component 
to aid the view change and state transfer mechanisms.
We implement the ability to instantiate \thesubsystem's multiple instances 
within a single trusted subsystem. 
Every instance can host a variable number of counters as needed by the protocol. 
For instance, Hybster requires certificates using at least three different 
counters for different protocol phases (e.g., checkpoints, view changes). 
Furthermore, the signing and the certifying functions must be hosted securely 
along with the private keys inside the trusted subsystem, 
while the signature aggregation and 
verification functions may be hosted outside as 
they only deal with public keys. 
We rely on attestation services provided 
by hardware vendors to verify that the secure code is running inside 
the enclave and perform any initialization steps.

\subsection{Linear Hybster}
\label{subsec:linear-hybster}

We now discuss the modifications to Hybster~\cite{behl2017hybrids}, 
to achieve linear communication in the common case
to create Linear Hybster. 

\begin{figure}[htbp]
	\centering
    \begin{subfigure}{0.49\textwidth}
        \centering
		\includegraphics[width=\linewidth]{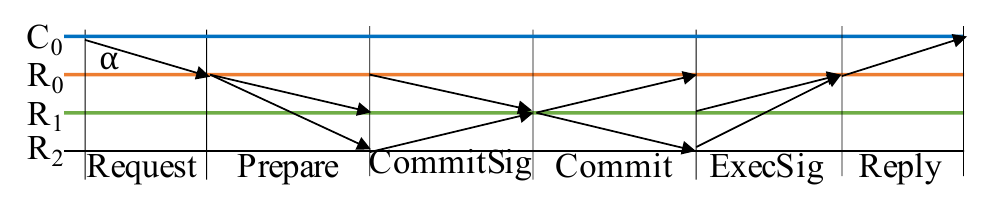}
        \vspace{-15pt}
		\caption{Linear Hybster}
        \label{fig:linhybster-example}
	\end{subfigure}
    \begin{subfigure}{0.49\textwidth}
        \centering
		\includegraphics[width=\linewidth]{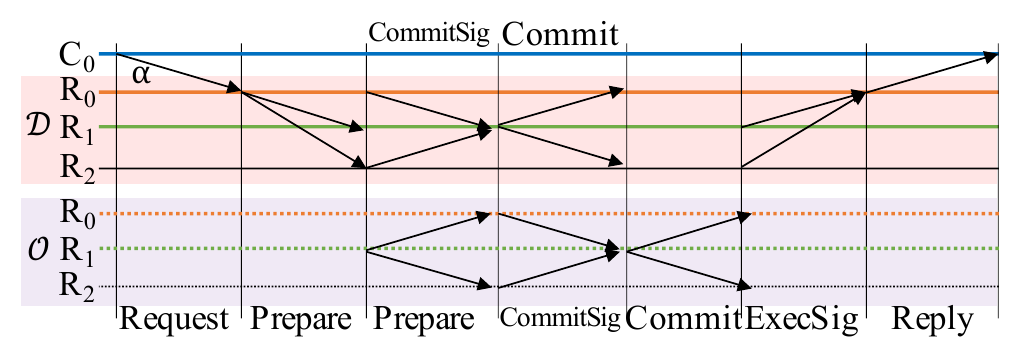}
        \vspace{-15pt}
		\caption{Destiny}
        \label{fig:destiny-example}
	\end{subfigure}
    \vspace{-10pt}
	\caption{Linear Hybster and \thesystem Agreement Protocol.}
\end{figure}

Figure~\ref{fig:linhybster-example} shows Linear Hybster's 
execution steps in the normal case. 
Hybster commits a command in two steps and requires clients to wait for $f+1$ 
replies. 
A quadratic number of \emph{Commit} messages are exchanged by replicas 
in an all-to-all communication, which bottlenecks throughput. 
We use a collector and an additional communication step to reduce 
this quadratic communication to linear. 
In \emph{Linear Hybster}, replicas send the \emph{Commit} messages to a 
collector (up to $f+1$ can be used for fault-tolerance), 
which aggregates at least $f+1$ messages and sends them to other replicas. 
Hybster uses \emph{TrInX}, a \emph{trusted} MAC provider 
which requires any pair of replicas to use unique secret keys to exchange 
messages between them. 
We replace \emph{TrInX} with \thesubsystem subsystem.  
\thesubsystem is configured with a threshold of 
$f+1$ out of $N = 2f+1$ total replicas. 

Hybster (and most BFT protocols) require that 
the clients wait for equivalent replies from at least $f+1$ replicas 
to defend against incorrect responses from malicious replicas. 
Linear Hybster, in contrast, reduces this $f+1$ communication to one 
single message using threshold signatures. 
For this purpose, Linear Hybster  uses another instance of 
threshold signatures, $\pi$, with threshold $f+1$. 
Now, once the client command is executed at each replica, the result of 
execution is signed using $\pi$ and sent to the collector in a 
\textsc{ExecSig} message. 
The collector collects and aggregates signature shares from $f+1$ 
valid \textsc{ExecSig} messages and generates an \textsc{ExecProof} message. 
This message is sent to the replicas as well as the client along with the 
result of execution. The client validates the aggregated signature, accepts the result, and returns.

\textbf{View Change}. A replica triggers a view change 
if it does not receive timely messages from the leader, or 
if it receives a proof that the leader is faulty 
(either via a publicly verifiable contradiction from the client 
or when $f + 1$ replicas complain).

Replica $R_m$ supports a new primary $R_p'$ of a view $v + 1$ 
by sending a \textsc{ViewChange} message with 
the \textsc{Prepare}s for all order numbers in its current ordering 
window in view $v$. 
A continuing counter certificate is attached to the message to ensure 
that even if replica $R_i$ is faulty, 
it includes all the \textsc{Prepare}s it is aware of up to 
the current order number. 
After sending the \textsc{ViewChange} message for $v + 1$, 
replica $R_m$ is prohibited from participating in view $v$. 
Due to the use of continuing counter certificates, 
a new leader $R_p'$ can determine all the proposals of 
the former primary $R_p$ by 
collecting only a quorum of \textsc{ViewChange} messages. 

Once a correct leader $R_p'$ collects 
at least $f+1$ \textsc{ViewChange} messages, 
it begins constructing the new view. It is possible that the new leader is 
lagging behind the current ordering window, 
in which case the new leader invokes the state-transfer protocol to 
request the checkpoint messages and the service state from an up-to-date replica. 
A replica cannot establish as a new leader until its ordering window matches with 
the \textsc{ViewChange} messages. 
Since only $f$ replicas can be faulty at most, 
there is at least one correct replica that contains the adequate 
information to help the new primary move to the new ordering window.

Unlike the agreement protocol, 
the view change mechanism uses continuing counter certificates 
provided by \thesubsystem. 
For a view change, 
replicas individually must announce their current view and their intended view, 
unlike normal case execution where replicas jointly accept a proposed command. 
Continuing counter certificates serve this purpose well, allowing replicas 
to individually prove their log state to other replicas and the new primary.

\subsection{Protocol}

\thesystem is obtained by instantiating \theparadigm using the 
Linear Hybster protocol presented in Section~\ref{subsec:linear-hybster}. 
\thesystem uses $7N$ messages and five phases in the optimistic case 
(seven in the pessimistic case) to execute each command 
(see Figure~\ref{fig:comparison-analysis-table}).
Due to linear communication, \thesystem's theoretical throughput 
closely matches the maximum concurrent throughput 
(Figure~\ref{fig:analytical-model-plot}).
For brevity, we provide an overview of \thesystem, leveraging 
the description in Section~\ref{sec:dqbft}.

\subsubsection{Agreement Protocol}
\thesystem commits both D-instances and O-instances strictly using the 
Linear Hybster protocol after being accepted by a majority, 
requiring a total of four communication steps. 
Note that the O-instance starts after the first communication step of the 
D-instance. Verification of execution results takes an additional two 
communication steps. 
The messages in the normal phase protocol are signed by invoking the 
\emph{Independent Counter Signature Shares} function of the corresponding 
\thesubsystem instance. 
This ensures the following properties: 
(i) \textit{Uniqueness}: the same counter value is not assigned to two 
different messages, and 
(ii) \textit{Monotonicity}: the counter value assigned to a message will 
always be greater than the previous counter value.

\subsubsection{Execution and Acknowledgement}
Replicas execute commands as they become \emph{ready} for execution.
After execution, as in Linear Hybster, 
replicas forward the signed result to a collector, which then aggregates 
$f+1$ signatures.
The collector sends this signature back to the replicas and to the client, 
indicating that the client's command was executed. 
Note that this step does not require the use of the trusted subsystem. 

\subsubsection{Checkpoint, State-transfer and View change Protocols}
\thesystem uses the respective checkpoint, state-transfer, and view change algorithms of the underlying Linear Hybster protocol. 

\begin{example}
    Figure~\ref{fig:destiny-example} illustrates with an example of 
    how \thesystem commits a command using the D- and O-instance protocols. 
    
    Assume that $R_1$ serves the primary role in the O-instance protocol.
    A client submits command $\alpha$ to replica $R_0$.
    $R_0$ becomes the \emph{initial} coordinator of $\alpha$. 
    We call it initial because if it fails,  some other replica will take over. 
    We also assume that $R_0$ and $R_1$ will play the collector roles for 
    the D-instance and O-instance, respectively. 
    A replica playing the collector role is responsible for 
    collecting signature shares, aggregating them, 
    and multicasting the combined signature.
    $R_0$ selects the lowest unused sequence number in its D-instance space, 
    assigns it to $\alpha$, 
    and disseminates the command by multicasting a \texttt{D-Prepare} message. 
    
    $R_1$ receives the \texttt{D-Prepare} message and 
    triggers the O-instance protocol for replica $R_0$. 
    $R_1$ proposes $R_0$'s ID and $\alpha$'s sequence number to 
    the next available O-instance sequence number (say $j$) and 
    sends a \texttt{O-Prepare} message.
    Replicas accept either Prepare messages and 
    send the corresponding \texttt{D-CommitSig} and \texttt{O-CommitSig} messages, 
    respectively, to the commit collectors, $R_0$ and $R_1$. 
    The \texttt{D-CommitSig} and the \texttt{O-CommitSig} messages are signed by 
    the $\sigma_0$ and $\tau$ \thesubsystem instances, respectively.
    The respective commit collectors wait for at least 
    $f+1$ valid \texttt{D-CommitSig} (respectively \texttt{O-CommitSig}) messages 
    and invoke the \thesubsystem subsystem to aggregate the signature shares in 
    the commit messages into a single signature. 
    The aggregated signatures are sent via the corresponding \texttt{D-Commit} 
    and \texttt{O-Commit} messages. 

    Replicas receive the valid \texttt{O-Commit} and \texttt{D-Commit} messages, 
    commit the command, and mark it for execution. 
    After executing the command at the global order number \emph{j}, 
    each replica  signs the resulting state and sends a signed 
    \texttt{ExecSig} message to the \emph{execution} collector. 
    The execution messages are signed using a \thesubsystem instance 
    that is different from the ones used during commit. 
    This \thesubsystem instance does not require trust and 
    is kept outside of the trusted subsystem.     
    The execution collector $R_0$ collects at least $f+1$ valid 
    \texttt{ExecSig} messages, aggregates the signatures into 
    a single \texttt{ExecProof} message, and sends the message to all the replicas. 
    It also sends this message to the client along with the result of execution. 
    The client verifies the signature, accepts the result, and returns.
\end{example}
\section{Evaluation}
\label{sec:eval}

We implemented multiple protocols under  
\theparadigm and evaluated them against state-of-the-art single-primary and 
multi-primary protocols.
Our evaluation answers the following questions:
\begin{compactenum}
	\item What is the impact of batching on protocol performance?
	\item How well do the protocols scale their performance when increasing 
	the system size from 10s to 100s of replicas in a geo-distributed 
	deployment?
	\item What is performance impact under replica failures?
	\item How do the \theparadigm protocols compare to other multi-primary protocols?
\end{compactenum}

\subsection{Protocols under test}

Our evaluation includes the following state-of-the-art protocols.

\subsubsection*{Single-primary protocols}
We evaluate PBFT~\cite{castro2002practical}, Hybster~\cite{behl2017hybrids}, 
and SBFT~\cite{Gutea:2019:SBFT}.
We use the variant of PBFT that  uses MACs that are computationally cheaper 
than signatures.
SBFT uses linear communication and $3f+c+1$ fast-path quorum with $3f+2c+1$ 
replicas. We set $c$ to zero, because increasing $c$ does not improve fault 
tolerance.
Chained Hotstuff~\cite{Yin:2019:HotStuff:3293611.3331591} is a rotating-primary 
protocol.

\subsubsection*{Multi-primary protocols}
Prime~\cite{Amir:2011:Prime:TDSC.2010.70} allows individual replicas 
to disseminate commands using Reliable Broadcast, and 
a primary provides an ordering for the disseminated commands periodically.
Dispel~\cite{Voron:2019:Dispel:arxiv:1912.10367} uses Reliable Broadcast 
to disseminate commands, and uses multiple instances of leaderless binary 
consensus to order the commands.
MirBFT~\cite{Stathakopoulou:2019:MirBFT:arxiv:1906.05552} 
allows multiple replicas to act as primaries concurrently 
by distributing sequence numbers evenly. 
It uses the notion of an epoch to define which replicas can be primaries 
during a certain period.
RCC~\cite{gupta2021rcc} allows multiple replicas to act as primaries and 
uses the notion of rounds to facilitate a global execution order. 
In each round, one command is committed by each of the primaries and 
a deterministic execution order is decided.

\subsubsection*{DQBFT protocols} 
DQPBFT, DQSBFT, and DQHybster are \theparadigm instantiations of the 
original protocols PBFT, SBFT, and Hybster, respectively. We also evaluated Linear Hybster and \thesystem.

We implemented all the protocols in a common framework in Golang. 
The framework uses \texttt{gRPC}~\cite{grpc} for communication and 
\texttt{protobuf}~\cite{protobuf} for message serialization. 
The ECDSA~\cite{Johnson:2001:ECD:2701775.2701951} algorithms 
were used for authenticating 
the messages exchanged by the clients and the replicas. 
We favored our own implementations over the author versions for  a fair and 
consistent evaluation. 
For instance, the authors' version of Hotstuff only disseminates 
command hashes~\cite{Stathakopoulou:2019:MirBFT:arxiv:1906.05552}, 
but all our implementations disseminate actual payloads.
In addition, the source code for RCC and Hybster were not 
publicly available.
The trusted components were implemented in C++ using the 
Intel SGX SDK~\cite{intel-sgx}.
Our implementations of BFT protocols perform out-of-order processing of 
commands, except Hotstuff, which does not support out-of-order processing 
because it rotates the primary's role regularly.
For Hybrid protocols, out-of-order processing is limited due to 
the use of counter-based trusted components.
Creation of signatures using the trusted components happen in order, 
whereas all other message processing happens out-of-order.

\subsection{Experimental Setup}

\begin{figure*}[!t]
	\centering
	\begin{subfigure}[b]{\textwidth}
		\centering
		\includegraphics[width=\textwidth]{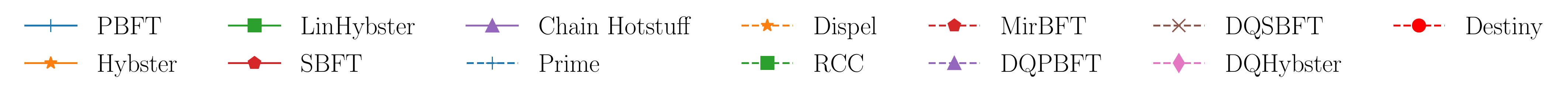}
	\end{subfigure}
	\begin{minipage}{.49\textwidth}
			\begin{subfigure}[b]{0.49\columnwidth}
				\centering
				\includegraphics[width=\textwidth]{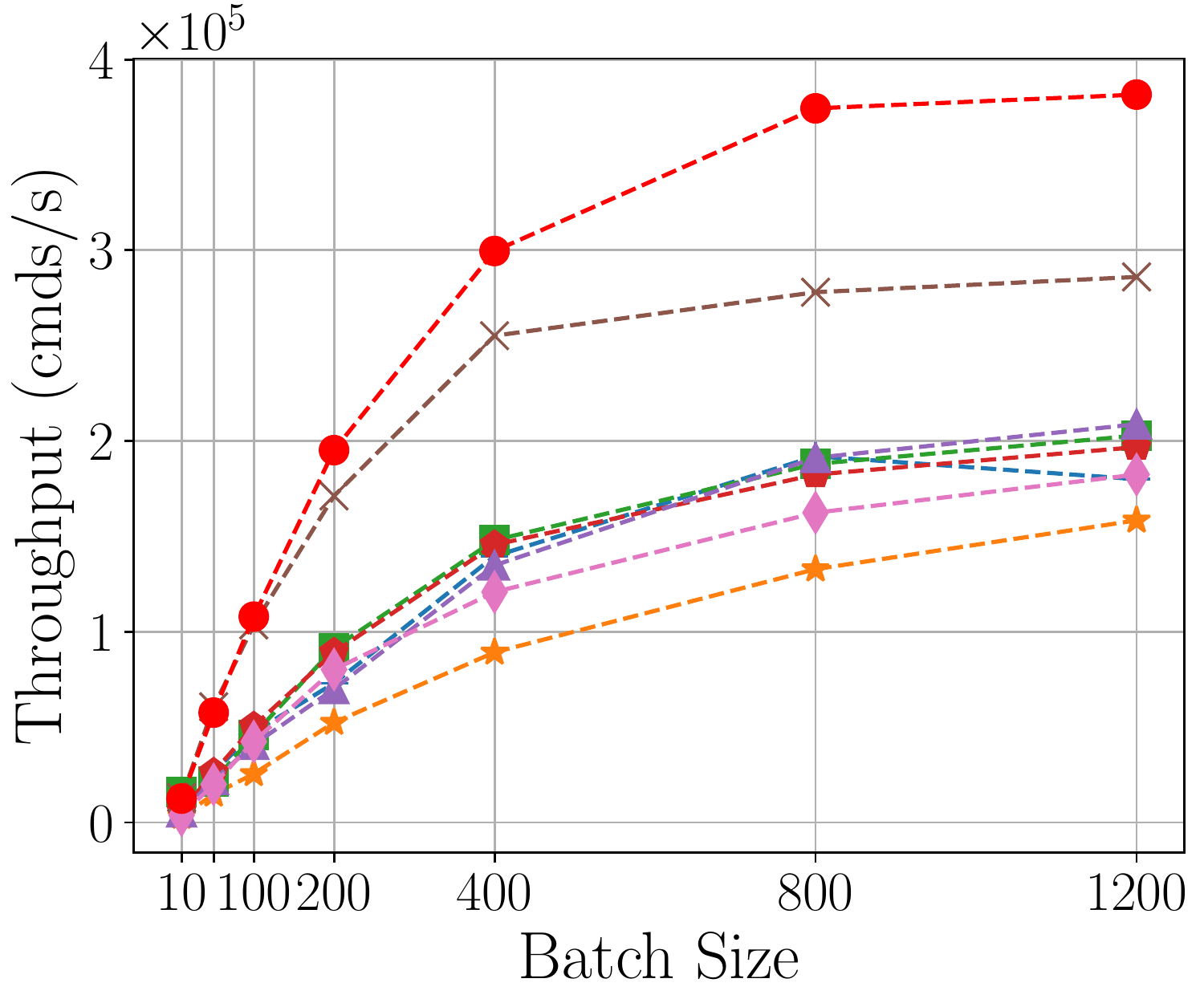}
				\caption{Multi-primary}
				\label{fig:mp-batching-tps}
			\end{subfigure}
			\begin{subfigure}[b]{0.49\columnwidth}
				\centering
				\includegraphics[width=\textwidth]{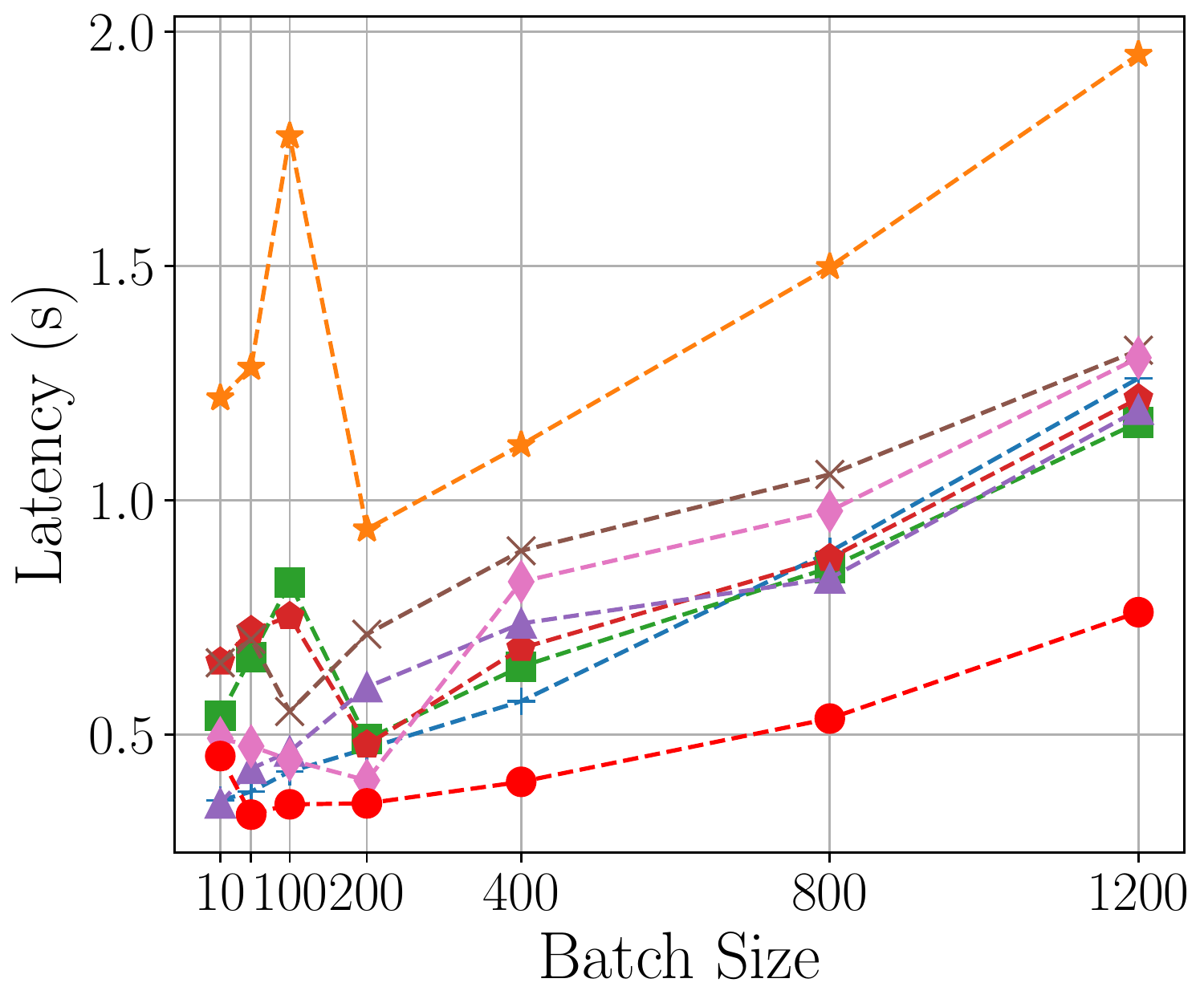}
				\caption{Multi-primary}
				\label{fig:mp-batching-lat}
			\end{subfigure}
			\begin{subfigure}[b]{0.49\columnwidth}
				\centering
				\includegraphics[width=\textwidth]{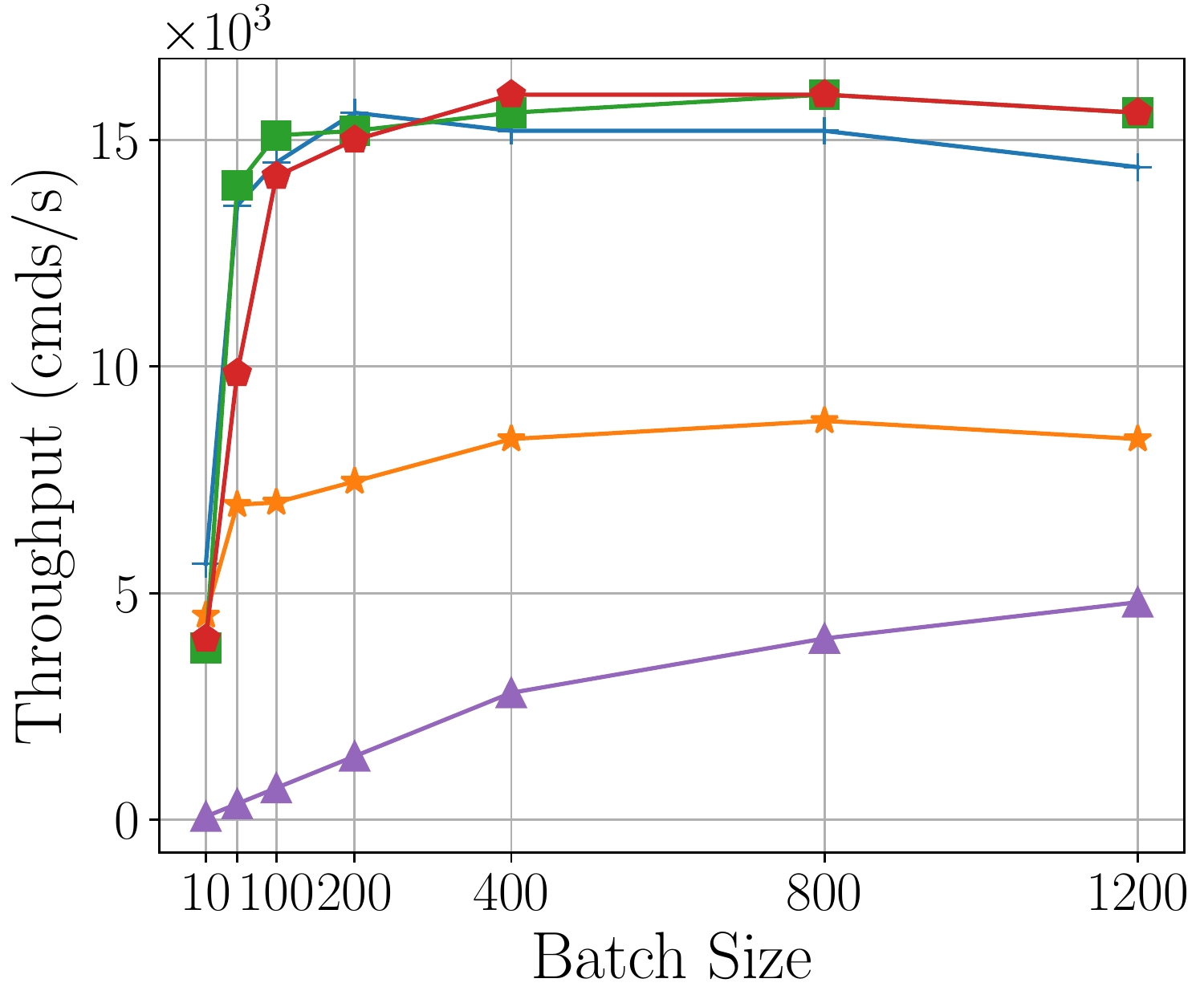}
				\caption{Single-primary}
				\label{fig:sp-batching-tps}
			\end{subfigure}
			\begin{subfigure}[b]{0.49\columnwidth}
				\centering
				\includegraphics[width=\textwidth]{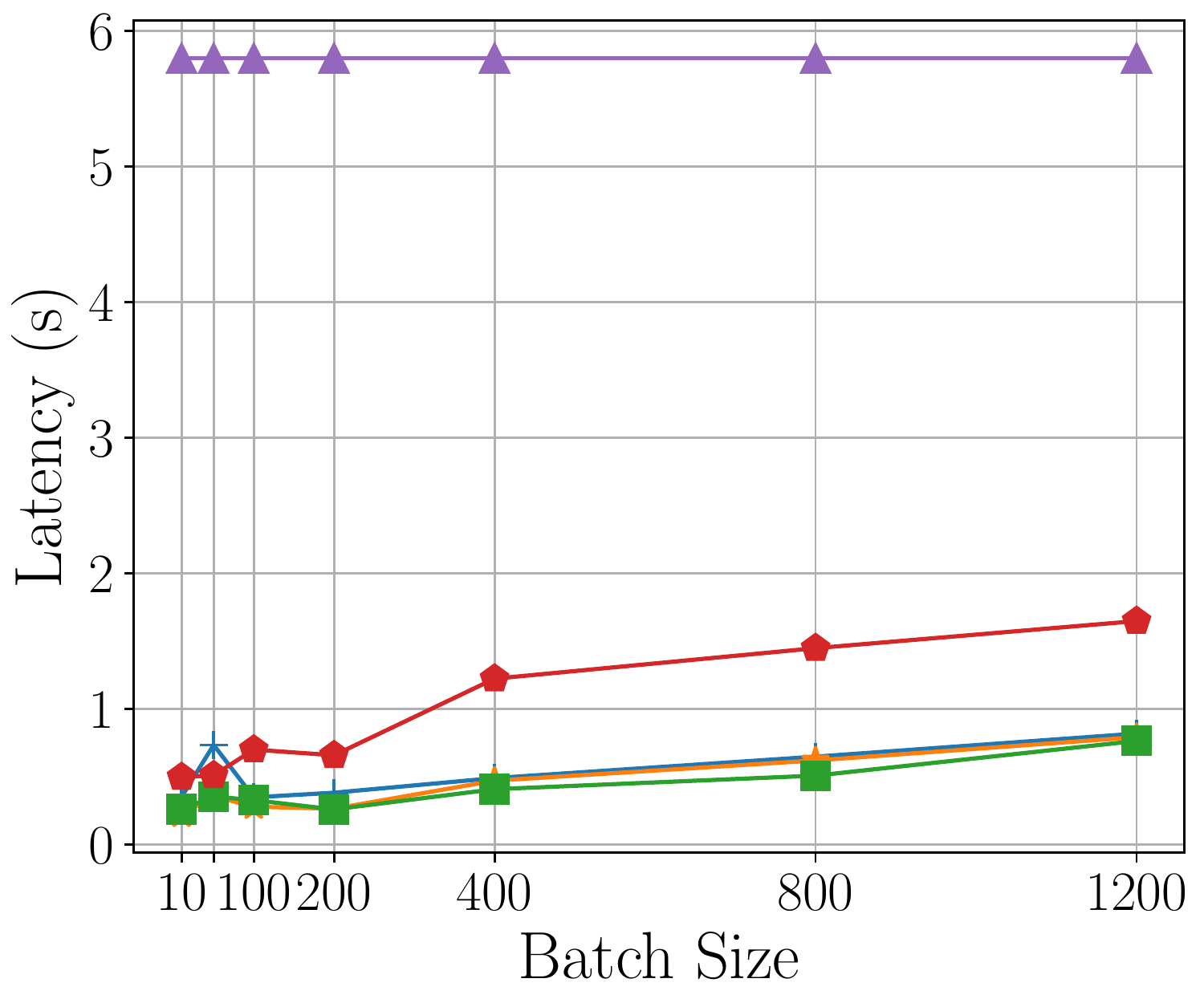}
				\caption{Single-primary}
				\label{fig:sp-batching-lat}
			\end{subfigure}
			\caption{Performance versus Batch Size (N=97).}
			\label{fig:batching-plot}
	\end{minipage}
	\begin{minipage}{.49\textwidth}
			\centering
			\begin{subfigure}[b]{0.49\columnwidth}
				\centering
				\includegraphics[width=\textwidth]{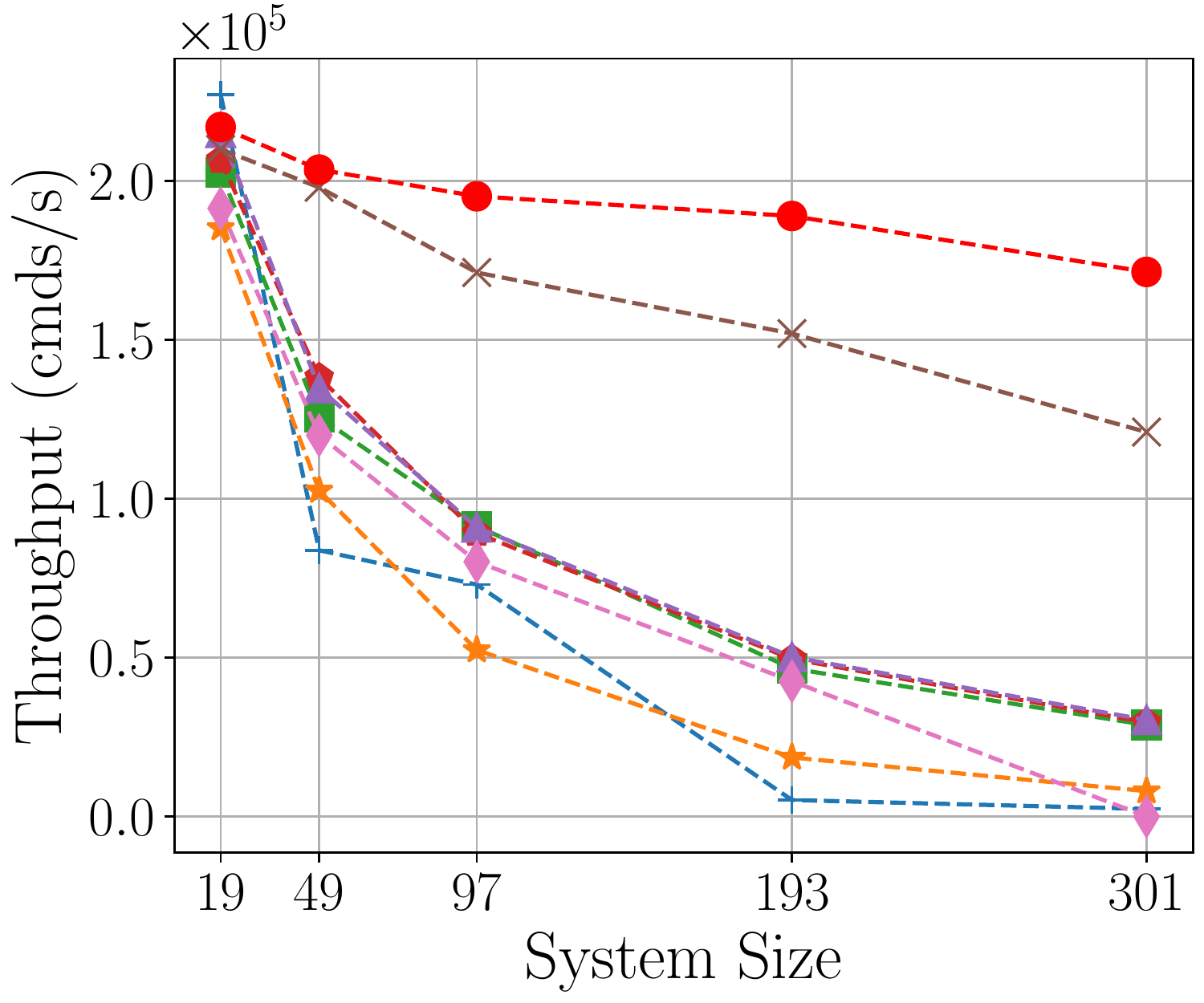}
				\caption{Scalability}
				\label{fig:mp-node-scalability-tps}
			\end{subfigure}
			\begin{subfigure}[b]{0.49\columnwidth}
				\centering
				\includegraphics[width=\textwidth]{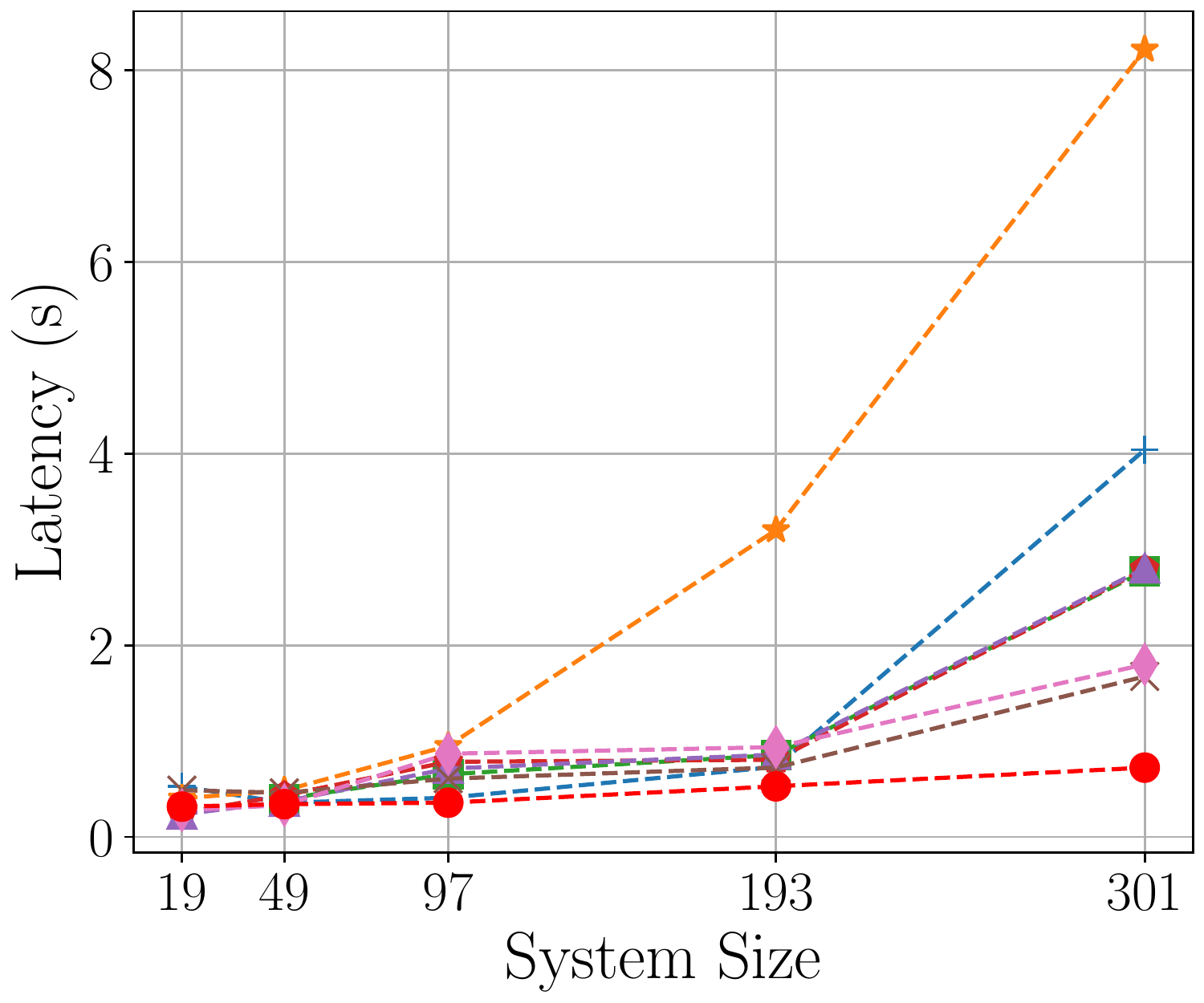}
				\caption{Scalability}
				\label{fig:mp-node-scalability-lat}
			\end{subfigure}
			\begin{subfigure}[b]{0.49\columnwidth}
				\centering
				\includegraphics[width=\textwidth]{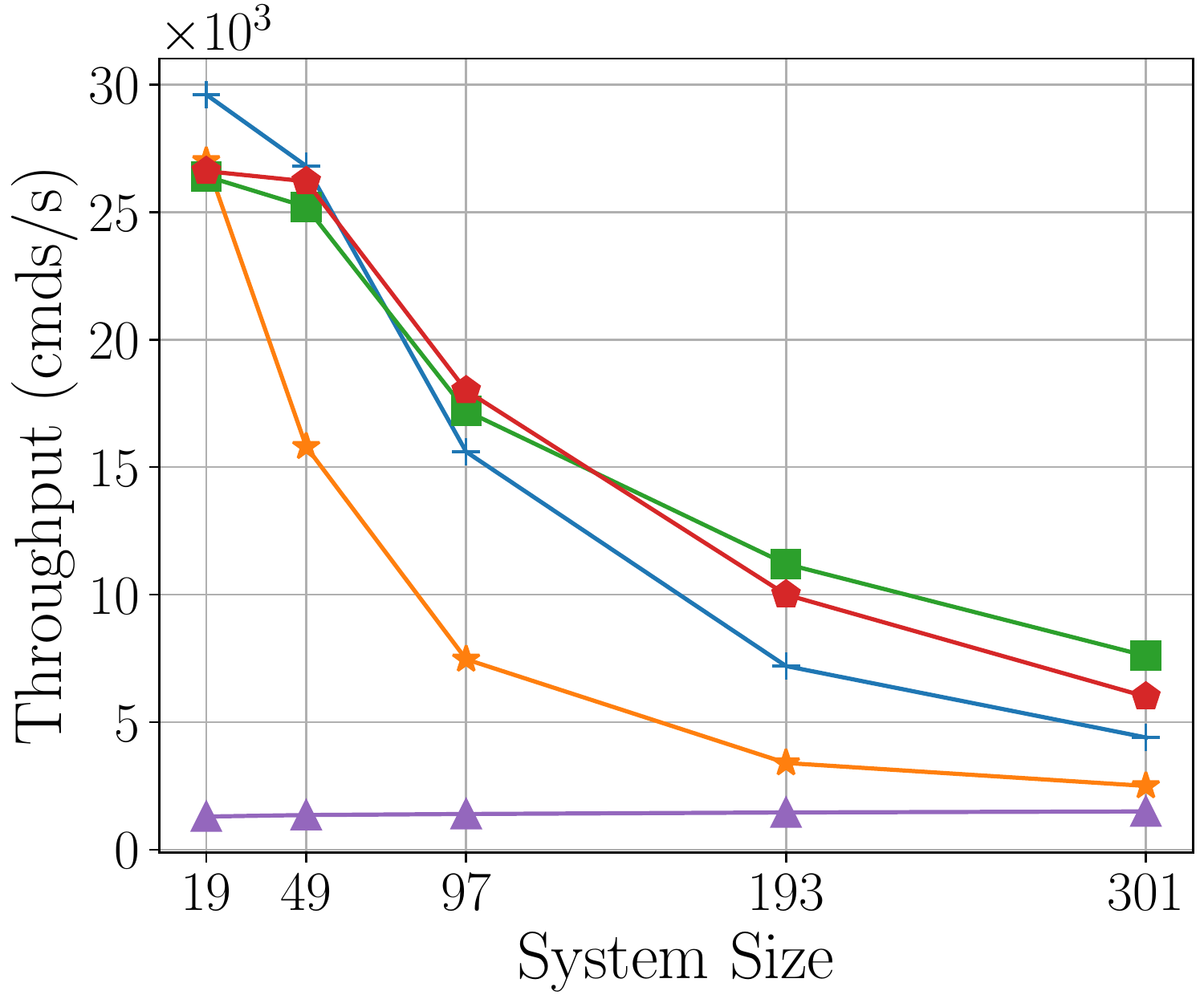}
				\caption{Scalability}
				\label{fig:sp-node-scalability-tps}
			\end{subfigure}
			\begin{subfigure}[b]{0.49\columnwidth}
				\centering
				\includegraphics[width=\textwidth]{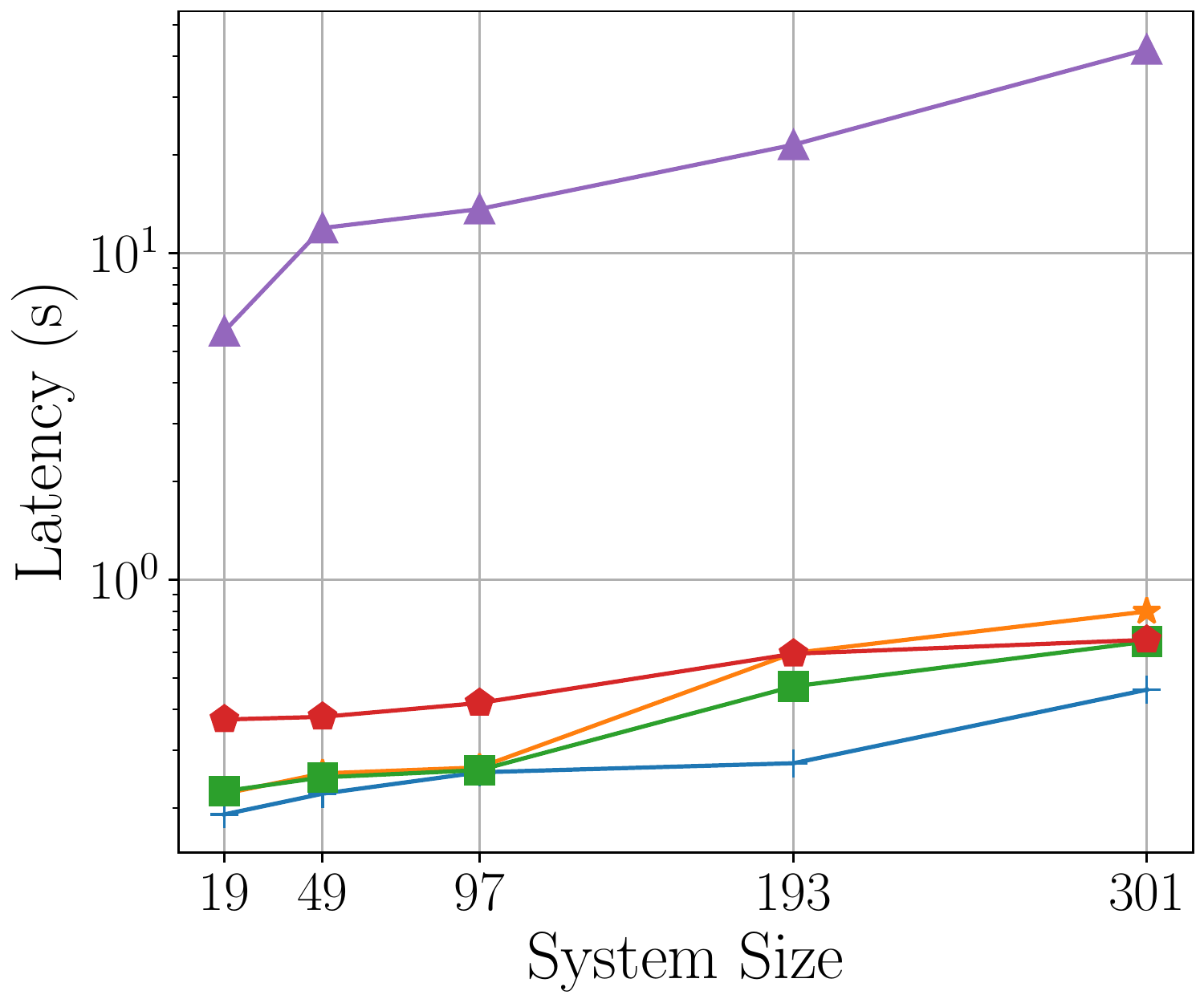}
				\caption{Scalability}
				\label{fig:sp-node-scalability-lat}
			\end{subfigure}
			\caption{Performance versus System Size.}
			\label{fig:scalability-plot}
	\end{minipage}
\end{figure*}

\begin{figure}[!h]
	\begin{subfigure}[b]{0.49\columnwidth}
		\centering
		\includegraphics[width=\textwidth]{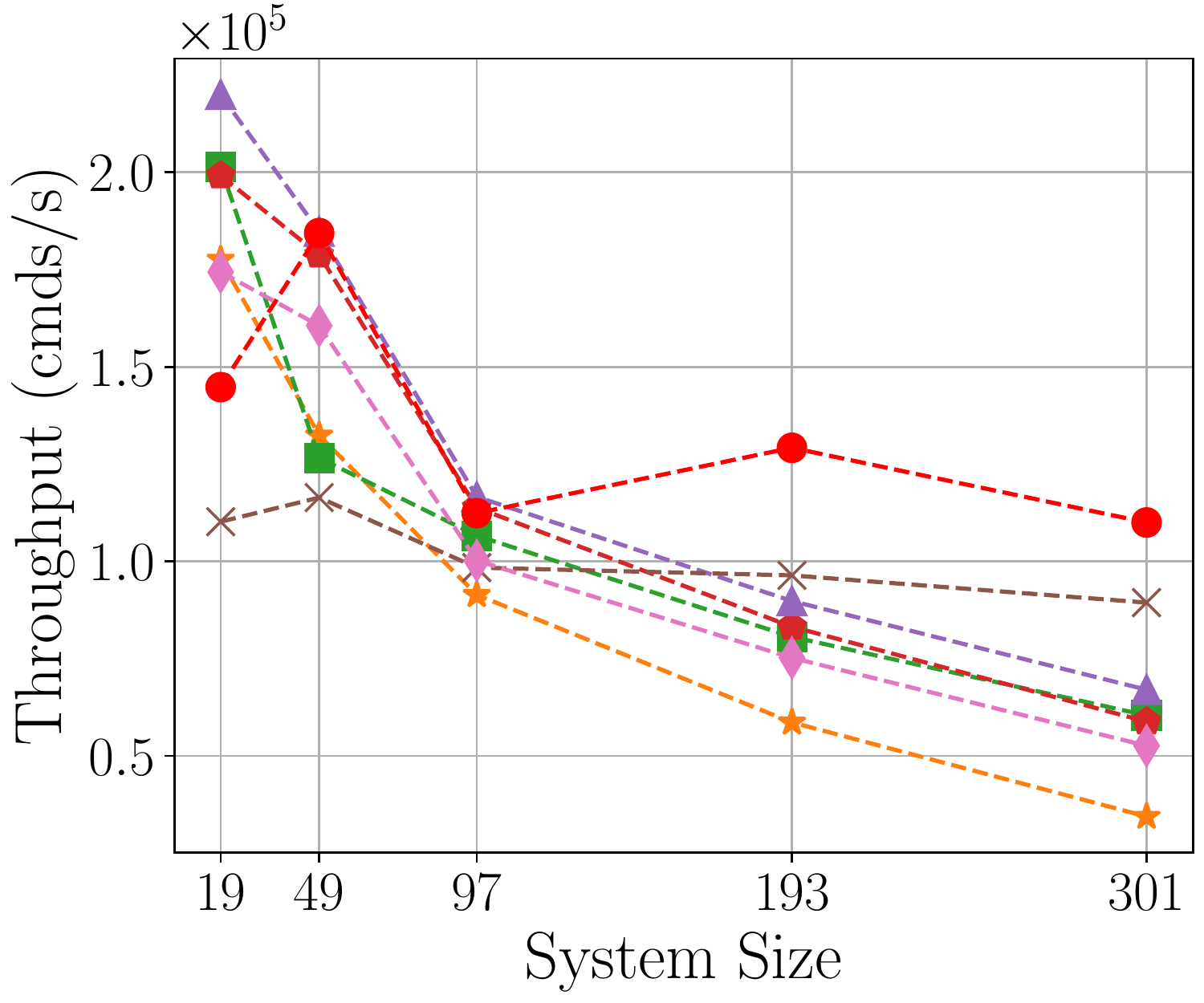}
		\caption{Multi-primary}
		\label{fig:mp-f-node-scalability-tps}
	\end{subfigure}
	\begin{subfigure}[b]{0.49\columnwidth}
		\centering
		\includegraphics[width=\textwidth]{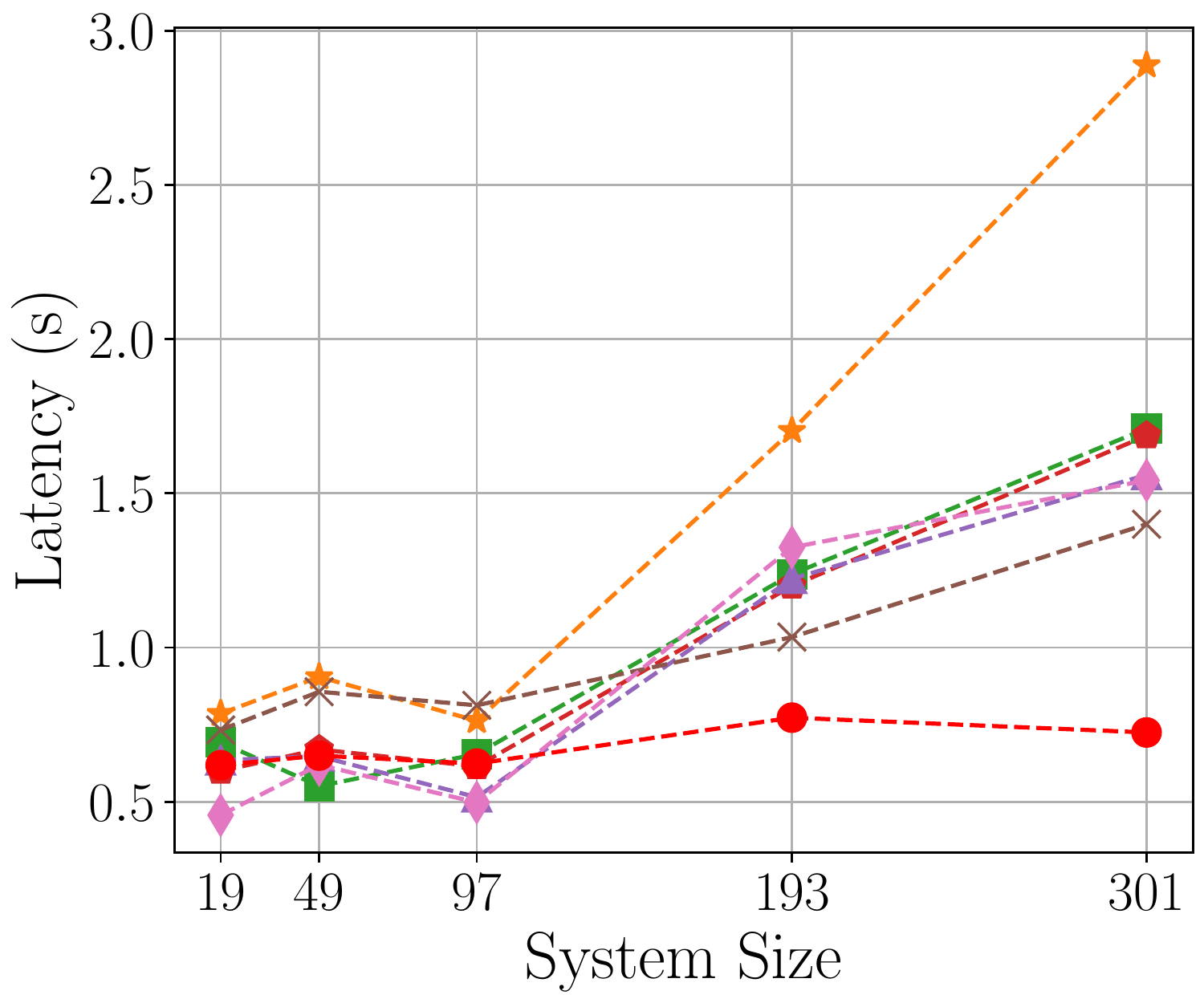}
		\caption{Multi-primary}
		\label{fig:mp-f-node-scalability-lat}
	\end{subfigure}
	\begin{subfigure}[b]{0.49\columnwidth}
		\centering
		\includegraphics[width=\textwidth]{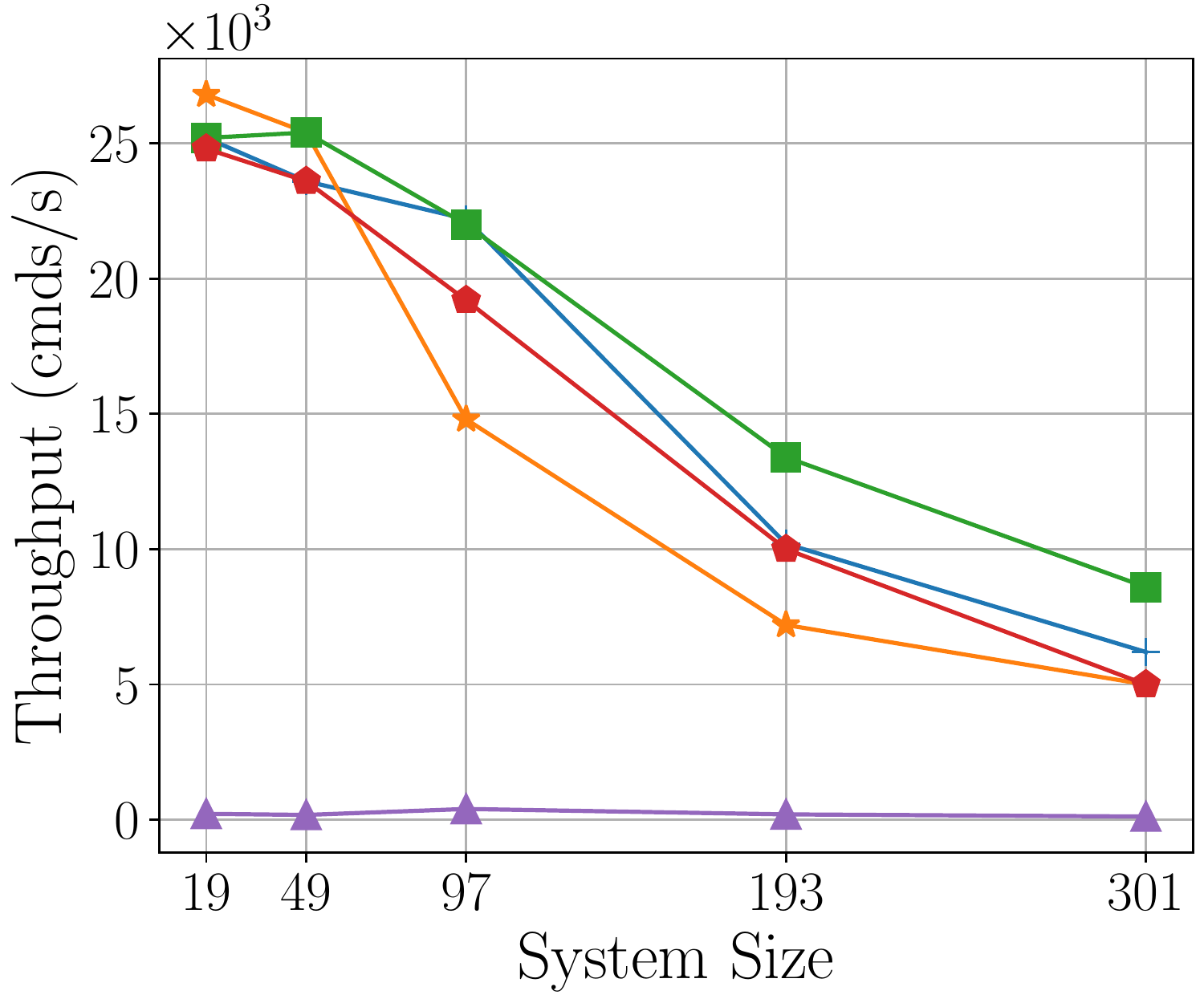}
		\caption{Single-primary}
		\label{fig:sp-f-node-scalability-tps}
	\end{subfigure}
	\begin{subfigure}[b]{0.49\columnwidth}
		\centering
		\includegraphics[width=\textwidth]{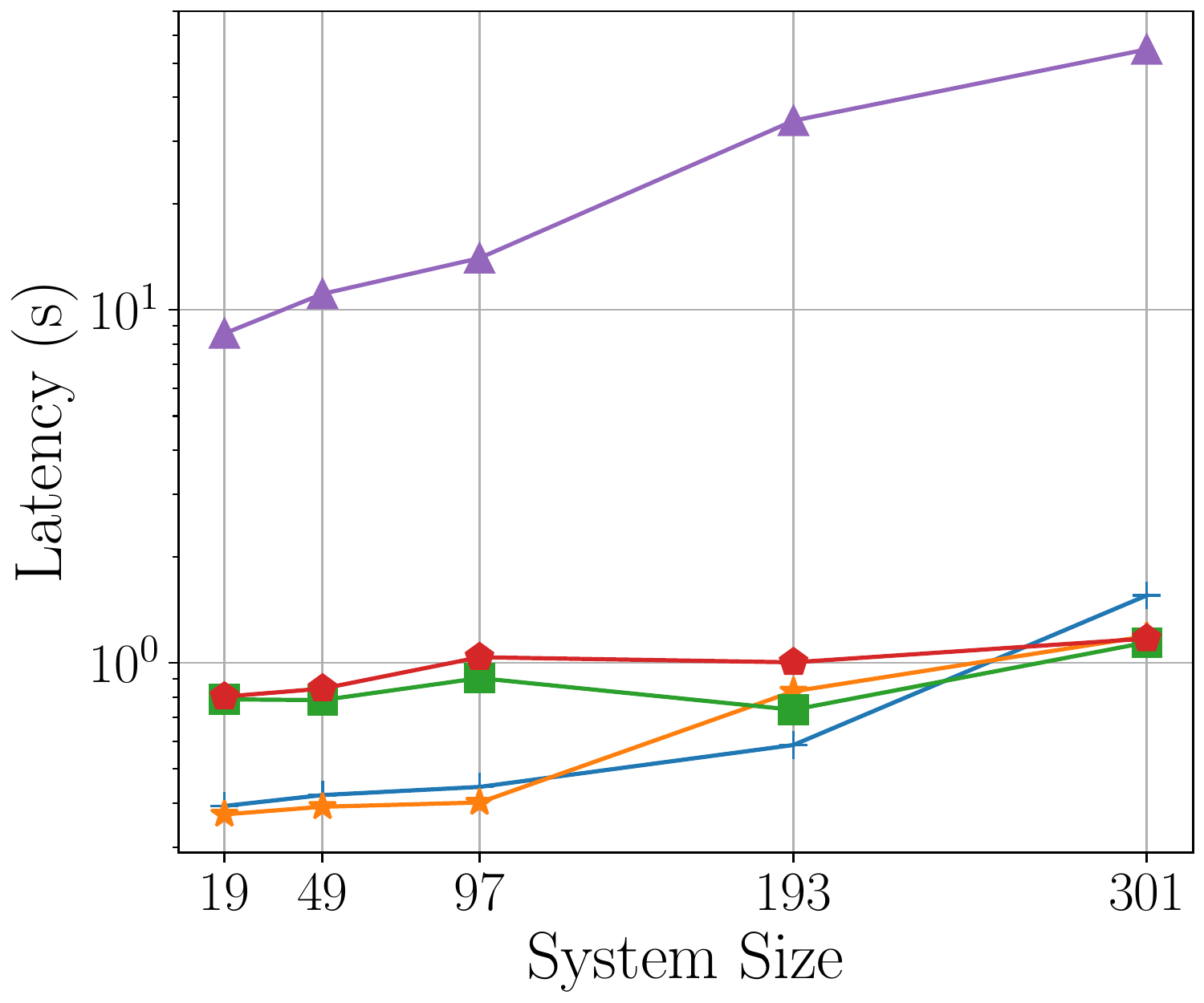}
		\caption{Single-primary}
		\label{fig:sp-f-node-scalability-lat}
	\end{subfigure}
	\caption{Performance under $f$ failures.}
	\label{fig:failures-scalability}
	\vspace{-12pt}
\end{figure}

We used SGX-enabled virtual machines (VMs)  available on Microsoft's Azure 
~\cite{azure-confidential-computing} platform. 
We obtained VMs from ten different datacenter regions: six in North America, 
three in Europe, and one in South East Asia. The protocols were deployed in 
each of these regions leveraging multiple VMs. 
The number of VMs depends on the experiment. 
Each VM consists of 8 vCPUs and 32GB of memory 
(best available at the time of experiments). 
The VMs were part of a 
Kubernetes~\cite{kubernetes} cluster and the protocol replicas and clients were 
deployed as \emph{pods}. 
We placed one replica pod per VM and placed the clients on different VMs. 
We designated a replica in Eastern US to serve the primary's role.
The network latencies between regions are in~\cite{azure-network-latency}.
The bandwidth between replicas ranged from 400 Mb/s (between US and Asia) and 
6 Gb/s (within same region).

We carried out experiments for five different values of $N$ 
(the number of replicas): 19, 49, 97, 193, 301 tolerating 6, 16, 32, 64, 100 
BFT and 9, 24, 48, 96, 150 Hybrid failures, respectively. 
For each experiment, replicas were evenly spread among the ten regions.
Clients send requests in a closed-loop, meaning they wait for the result of the 
previous request before sending the next one. 
Unless otherwise stated, 
clients are evenly spread across all the region and send commands to their 
local replicas for multi-primary protocols and to the primary for 
single-primary protocols.
Our performance numbers account for both the consensus and execution 
time.
We use Prometheus~\cite{208864} timeseries database to collect metrics 
from the replicas \emph{periodically} and report our results.
The state is a fully-replicated in-memory key value store, 
a useful abstraction for building other applications 
including smart contract engines~\cite{Gutea:2019:SBFT}. 
The workload is 100\% put operations with 20-byte keys and random values.
The command payload size was set at 512 bytes. 
Unless otherwise stated, we use a batch size of 200 client commands per batch.

\subsection{Experiments}

\subsubsection{Batching Experiment}
\label{sec:eval:batching}

First, we measured the impact of batching commands 
on protocol performance. 
Increasing the batch size increases the size of the initial phase message
multicasted by the primary (or the coordinator in the case of \theparadigm).
For this experiment, we deployed $N=97$ replicas, increased the batch size 
from 10 to 1200 commands per batch, and measured the performance. 
Figure~\ref{fig:batching-plot} shows the results.
The single-primary protocols reach their maximum throughput at batch size of 
100, as their primaries' are saturated. 
Hybster's performance is limited because it performs 
in-order attestation of messages including those with command 
payloads by copying them into the trusted component hosted inside the enclave.
In contrast, linear communication complexity and the use of threshold 
signatures in Linear Hybster pays off as only the command hash is copied to 
the enclave enabling it to perform at par with PBFT that uses cheaper 
message authentication codes (MAC).
Moreover, threshold signatures also help SBFT and Linear Hybster scale 
better at large batch sizes than MAC.
Chained Hotstuff's 
throughput is significantly limited because it rotates the 
primary for each command that disallows out-of-order processing of multiple 
batches simultaneously. 
Thus, its latency is higher because each replica must wait for 96 
other replicas to propose before its turn.

The multi-primary protocols show multifold increase in throughput 
compared to single-primary protocols by virtue of allowing multiple replicas 
to propose simultaneously.
RCC, MirBFT, and DQPBFT perform similarly because under non-faulty 
scenarios their effective behaviors are the same. 
Note that RCC is also a BFT paradigm 
and can also be instantiated with SBFT; 
we observed that its performance to be on par with DQSBFT's performance 
given that the load.
\thesystem's performance exceeds all other protocols with 35\% better 
throughput than the next best protocol DQSBFT and 40\% lower latency than 
other multi-primary protocols. 
\thesystem performs better because aggregating $f+1$ signature shares 
is computationally cheaper than aggregating $3f+1$ shares
~\cite{Tomescu:2020:ScalableThreshold:SP40000.2020.00059}, 
and the $f+1$ quorum gives $f$ additional replicas to provide 
redundancy from slow nodes and staggering network, unlike SBFT.
Note that this experiment also serves to demonstrate the impact of increasing 
the command size because the execution overheads are small for our key-value 
store. 
For instance, a command payload of 1024 bytes and 200 batch size will have 
similar performance to a command payload of 512 bytes and 400 batch size.

\subsubsection{Scalability Experiment}
\label{sec:eval:scalability}

Second, we measured the performances of the protocols while increasing the 
system size, i.e., the number of replicas, from 19 to 301 replicas. 
Figure~\ref{fig:scalability-plot} shows the results.
Similar to the previous experiment, 
the performance of single-primary protocols is limited by their primaries' 
bandwidth. 
Thus, their performances decrease with increasing $N$ since the
primaries must send the initial payload ($\approx100kB$) to all the replicas.

On the other hand, multi-primary protocols have a higher peak throughput 
than single-primary 
protocols by virtue of enabling multiple replicas to send the initial payload 
that distributes the bandwidth requirements among all replicas.
As with the batching experiment, the performance trends for 
RCC, MirBFT, and DQPBFT are similar. 
\thesystem's throughput scales better than all other protocols.
At 301 replicas, \thesystem's provides $40\%$ better throughput and 70\% 
lower latency than the next best protocol DQSBFT.

We also analyzed the CPU utilization and network traffic at $N=97$.
For single-primary protocols, the primaries reached peak traffic of 
6Gbps at CPU usage between 50\%-65\%, while the replicas bandwidth was 
$\approx$115Mbps with 10\%-20\% CPU utilization.
For \theparadigm protocols, the average replica traffic was $\approx$1.5Gbps 
and CPU usage was 65\%. 
\thesystem reached peak CPU usage of 95\% indicating that the other
\theparadigm protocols were limited by their bandwidth 
(inline with Figure~\ref{fig:analytical-model-plot}).

\subsubsection{Scalability under Failures}

Third, we measured the protocol performance under failures. 
For this purpose, we repeated the scalability 
experiment with $f$ failed replicas.
Failed replicas are equally spread between the ten regions 
similar to the deployment spread.
Figure~\ref{fig:failures-scalability} shows the results.
Note that SBFT, DQSBFT, and \thesystem are more negatively impacted by the 
failure of $f$ replicas than $DQPBFT$, RCC, and MirBFT. 
Both SBFT and $DQSBFT$ must fallback to the slow path by default since they 
lack the fast quorum, which is equal to the system size. 
This adds additional communication steps to the SBFT protocols.
Thus, the performance of DQSBFT is poor and not any better than DQPBFT.
On the other hand, \thesystem must wait for replies from all the regions 
instead of only from a majority of regions, because the majority of regions do not 
have the majority quorum due to failures.
Yet, \thesystem performs better than others at 193 and 301 replicas 
because its linear communication pays off at that scale.

Note that due to the reduction in the number of replicas 
that participate in a given round, Dispel's latency under failures is 
substantially lower than that during the failure-free case. 
Similarly, for other multi-primary protocols, we observe lower latencies than 
with failure-free experiments. 
We attribute this to the reduction in number of messages that are 
sent and processed by each replica.

\subsubsection{Single Replica Failure Experiments}

Although the previous experiments show that \thesystem performs better than state-of-the-art multi-primary protocols such as RCC, the \theparadigm protocols (DQSBFT and DQPBFT) perform only at par with the
RCC paradigm protocols. 
In the previous experiments, we balanced the clients equally on all the 
regions and ensured that the replicas receive requests from the clients at the 
same rate. 
However, in practice, it may not be feasible to ensure a uniform request rate 
among replicas, because certain regions may have more load than others, 
e.g., due to geographical characteristics such as different time zones, 
or even Byzantine behaviors.
Therefore, we devised an experiment to compare the performance of DQPBFT 
with other multi-primary protocols, RCC and MirBFT, when different replicas 
receive requests at different rates.

\begin{figure}[!t]
	\includegraphics[width=0.85\columnwidth]{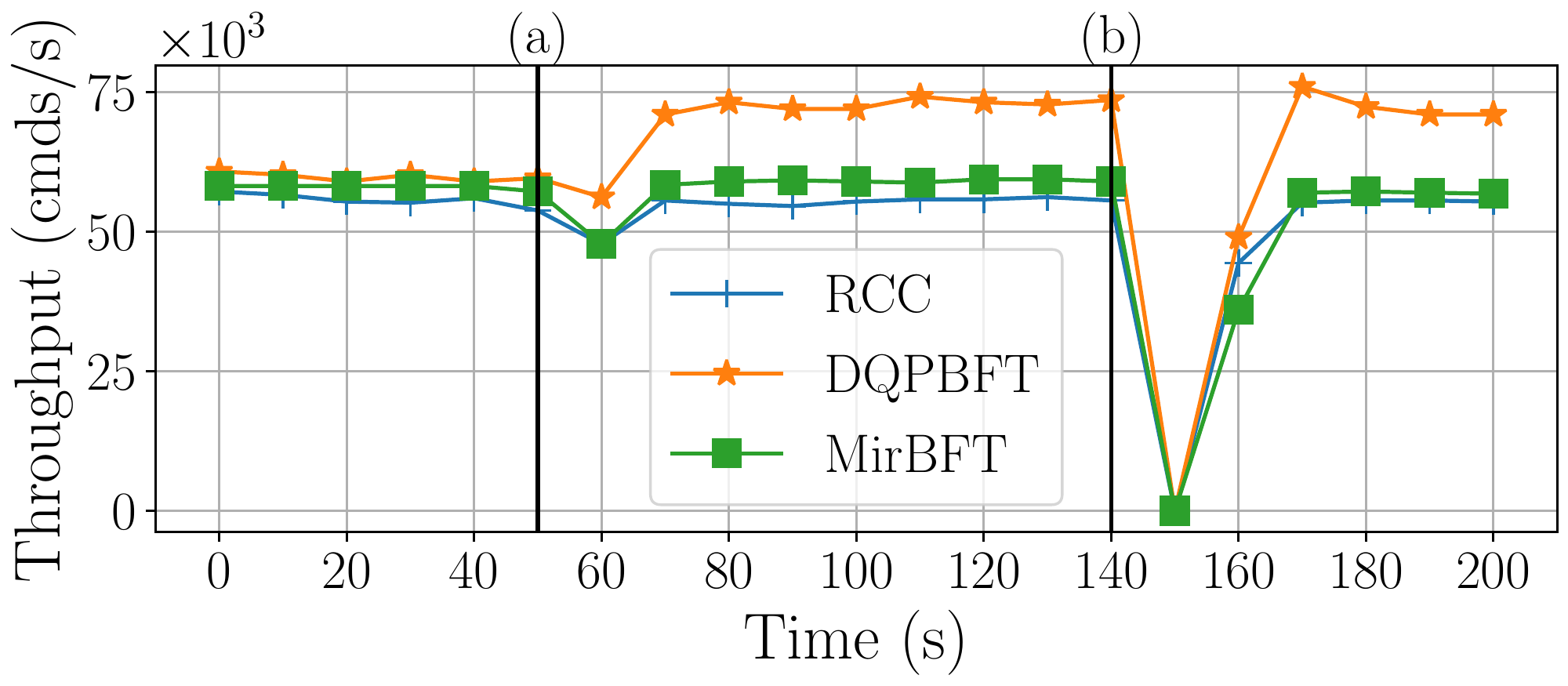}
	\caption{Throughput timeline with slow replicas and injected failures.
	At (a), the number of clients is doubled in all but one region 
	causing a replica to slow down. At (b), a random replica 
	is killed to invoke the view change procedure.}
	\label{fig:failure-timeline}
\end{figure}

For this experiment, we deployed 97 replicas that are spread among the ten regions,  
and increased the number of clients non-uniformly over time. Figure~\ref{fig:failure-timeline} shows the results. Initially, at $t=0$, clients are spread evenly among the replicas, 
during which all three protocols, namely, RCC, DQPBFT, and MirBFT, 
perform similarly as their behavior is effectively the same under these 
conditions. At $t=50$, we double the number of clients in all regions except in one 
region, namely South East Asia. 
Following the increase, at $t=60$, as the replicas 
are overwhelmed by the sudden increase in requests from the new clients, 
they lose throughput slightly for some time before bouncing back. 
As the system stabilizes, it can be observed that DQPBFT's throughput 
increases by 25\% while that of RCC and MirBFT remain the same as before. 
The O-instance of the DQBFT paradigm enables each replica to 
deliver commands at its own pace without depending on other replicas' deliveries.
In contrast, the round-robin approach to delivery used by MirBFT and RCC 
throttles all the replicas to deliver commands at the rate of the slowest 
replica.

When a replica becomes faulty, all three protocols cease to 
deliver client requests because the undelivered commands from the faulty 
replica must be delivered. Figure~\ref{fig:failure-timeline} shows this effect 
at $t=140$, when one replica is killed. All three protocols cease to 
deliver commands as they begin their view change procedures for the failed 
replica. Once the view change completes, throughput is restored.
\section{Related Work}
\label{sec:relwork}

Numerous performance-oriented single-primary BFT protocols
~\cite{kotla2007zyzzyva,Gutea:2019:SBFT,aublin2015next,castro2002practical,veronese2013efficient} 
 have been proposed in literature.
In Section~\ref{sec:bg:dc}, we discussed the limitations of primary-based, 
the rotating-leader~\cite{Yin:2019:HotStuff:3293611.3331591,Veronese:2009:SOW:1637865.1638341,Veronese:2010:EEB:1909626.1909800} 
and dependency-based ordering~\cite{ezbft,guerraoui2010next} 
approaches.

Request dissemination~\cite{Amir:2011:Prime:TDSC.2010.70,10.5555/866693,10.1145/1774088.1774187,Clement:2009:UCS:1629575.1629602} 
has been proposed as a means to relieve primary's workload. 
These solutions use Reliable Broadcast that increases the overall latency 
since replication must always precede ordering due to lack of 
Agreement property and 
incurs quadratic communication. 

The idea of separating replication and global ordering has been explored 
in the crash fault 
model~\cite{Zhao:2018:SDPaxos:3267809.3267837,Li:2016:NoPaxos:OSDI:3026877.3026914,Balakrishnan:2013:Corfu:2535930}. 
SDPaxos~\cite{Zhao:2018:SDPaxos:3267809.3267837} separates replication from 
ordering, and uses a consensus protocol for both the tasks. 
\theparadigm's separation technique can be viewed as the BFT 
counterpart, but our design is optimized for scalability to 
hundreds of replicas, while SDPaxos focuses on minimizing latency in up to
five-replica deployments.
Furthermore, distributed log protocols 
(e.g. Corfu~\cite{Balakrishnan:2013:Corfu:2535930}) use
a benign sequencing node to dictate global order.
To prevent malicious sequencers from violating consistency, 
the O-instance in \theparadigm must assign sequence numbers by 
reaching BFT consensus.
Moreover, the interaction between D- and O- instances must ensure that 
none of the instances compromise the safety/liveness properties of each other.
NoPaxos~\cite{Li:2016:NoPaxos:OSDI:3026877.3026914} requires special network 
devices, thus is suitable only within a datacenter. 

Various trusted component designs have been proposed previously for Hybrid 
protocols~\cite{behl2017hybrids,a2m,veronese2013efficient,kapitza2012cheapbft}. 
The trusted counter design is simple and memory-efficient 
compared to log-based designs. Among the known Hybrid protocols, we chose 
Hybster because the protocol's is designed specifically for commodity 
processors with trusted subsystems such as Intel SGX.
Threshold secret shares can be used in 
place of threshold signatures~\cite{Liu:2019:FastBFT}, but
requires creating a set of secret key shares for 
each command and exposing it when committing each command. 
This requires additional computational and network resources.
PoET~\cite{sawtooth} uses the trusted component to dictate the minimum time 
period between block proposals, thus adopting the synchrony timing model.

Alternate (e.g. XFT~\cite{liu2016xft}) and 
mixed fault models (e.g. Hierarchical~\cite{Gupta:2020:ResilientDB:3380750.3380757,Amir:2010:Steward:TDSC.2008.53}) 
have been proposed to improve performance in geo-distributed systems. 
XFT assumes synchronous communication among majority replicas for safety, 
while the Hybrid protocols assumes the trusted component for safety.
Unlike mixed fault models, it tolerates $f$ global failures and has no limits on 
regional failures.   
\section{Conclusion}
\label{sec:conclusion}

In conclusion, 
we show that \theparadigm is an effective paradigm for designing 
highly scalable BFT protocols. 
Furthermore, with \thesystem, we show that linear communication and smaller 
quorums elevate the performance of \theparadigm protocols.


\bibliographystyle{ACM-Reference-Format}
\bibliography{bft-refs,destiny-refs}

\end{document}